\documentclass[aps,prd,preprint,endfloats,showpacs,superscriptaddress,nofootinbib]{revtex4}
\usepackage{amsmath,amssymb,amsbsy,color}
\usepackage{moreverb}
\usepackage[dvips]{graphicx}
\usepackage{psfrag}
\usepackage{indentfirst}

\begin{document}

\draft
\preprint{UTHEP-533}
\preprint{UTCCS-P-29}

\title{Neutron electric dipole moment with external electric field method in lattice QCD}
\author{E.~Shintani}
\affiliation{
Graduate School of Pure and Applied Sciences, University of Tsukuba,
Tsukuba, Ibaraki 305-8571, Japan }

\author{S.~Aoki}
\affiliation{
Graduate School of Pure and Applied Sciences, University of Tsukuba,
Tsukuba, Ibaraki 305-8571, Japan }
\affiliation{
Riken BNL Research Center, Brookhaven National Laboratory, Upton, 11973,
USA}

\author{N.~Ishizuka}
\affiliation{
Graduate School of Pure and Applied Sciences, University of Tsukuba,
Tsukuba, Ibaraki 305-8571, Japan }
\affiliation{
Center for Computational Sciences, University of Tsukuba, Tsukuba,
Ibaraki 305-8577, Japan }

\author{K.~Kanaya}
\affiliation{
Graduate School of Pure and Applied Sciences, University of Tsukuba,
Tsukuba, Ibaraki 305-8571, Japan }

\author{Y.~Kikukawa}
\affiliation{Institute of Physics, University of Tokyo, Komaba,
Tokyo 153-8902, Japan}

\author{Y.~Kuramashi}
\affiliation{
Graduate School of Pure and Applied Sciences, University of Tsukuba,
Tsukuba, Ibaraki 305-8571, Japan }
\affiliation{
Center for Computational Sciences, University of Tsukuba, Tsukuba,
Ibaraki 305-8577, Japan }

\author{M.~Okawa}
\affiliation{
Department of Physics, Hiroshima University, Higashi-Hiroshima,
Hiroshima 739-8526, Japan }

\author{A.~Ukawa}
\affiliation{
Graduate School of Pure and Applied Sciences, University of Tsukuba,
Tsukuba, Ibaraki 305-8571, Japan }
\affiliation{
Center for Computational Sciences, University of Tsukuba, Tsukuba,
Ibaraki 305-8577, Japan }

\author{T.~Yoshi\'e}
\affiliation{
Graduate School of Pure and Applied Sciences, University of Tsukuba,
Tsukuba, Ibaraki 305-8571, Japan }
\affiliation{
Center for Computational Sciences, University of Tsukuba, Tsukuba,
Ibaraki 305-8577, Japan }

\collaboration{ CP-PACS Collaboration }

\date{\today}

\begin{abstract}
We discuss a possibility that the Neutron Electric Dipole Moment (NEDM) 
can be calculated in lattice QCD simulations in the presence of the CP 
violating $\theta$ term.
In this paper we measure the energy difference between spin-up and
spin-down states of the neutron in the presence of an uniform and
static external electric field.
We first test this method in quenched QCD with the RG improved gauge
action on a $16^3\times 32$ lattice at 
$a^{-1}\simeq$ 2 GeV, employing two different lattice fermion
formulations, the domain-wall fermion and the clover fermion for
quarks, at relatively heavy quark mass $(m_{PS}/m_V \simeq 0.85)$.
We obtain non-zero values of NEDM from calculations with both fermion 
formulations. 
We next consider some systematic uncertainties of our method for
NEDM, using $24^3\times 32$ lattice at the same lattice spacing only with the
clover fermion.
We finally investigate the quark mass dependence of NEDM and 
observe a non-vanishing behavior of NEDM toward the chiral limit.
We interpret this behavior as a manifestation of the pathology in the 
quenched approximation.
\end{abstract}


\maketitle


\section{Introduction}
Discrete symmetries, such as parity (P), charge conjugation (C) and 
time-reversal (T),  have played important roles to establish the structure of 
the standard model. One of the most famous examples is CP violation 
which led to three generations of quarks and 
leptons\cite{KM}. 

In the strong interaction, the most strict constraint on violation of P and 
T symmetries comes from the measurement of the electric dipole moment (EDM) 
for neutron (NEDM) and proton (PEDM).
The current upper bound is given by
\begin{equation}\label{exp_dn}
  |d_N| < 6.3\times 10^{-13}\,\,e\cdot\textrm{fm}\,\textrm{(90\%C.L.)}\,
\end{equation}
for neutron from a Larmor frequency 
measurement with ultra-cold neutron (UCN)\cite{Harris}, 
and
\begin{equation}
  |d_N| < 5.4 \times 10^{-11}\,\,e\cdot\textrm{fm}
\end{equation}
for proton\cite{Dmitriev},
which is estimated indirectly from EDM of mercury atom 
${}^{199}\rm{H}_{\rm{g}}$ given by
$d_{\rm atom}({}^{199}\rm{H}_{\rm{g}})<2.1\times 10^{-15}\,e\cdot
\rm{fm}\,\textrm{(95\% C.L.)}$\cite{Romalis}.

On the other hand, QCD allows a gauge invariant, renormalizable, and CP odd 
$\theta$ term,
\begin{eqnarray}
&& i\frac{\theta}{32\pi^2}\int d^4x\,\widetilde G_{\mu\nu}(x)G_{\mu\nu}(x)
   \label{L_QCD}, \qquad
   \widetilde G_{\mu\nu}(x) = \frac{1}{4}\epsilon_{\mu\nu\alpha\beta}
G_{\alpha\beta}(x)
\end{eqnarray}
in Euclidean space-time
with $G_{\mu\nu}$ which is  the field strength of gluon.
Some model estimations\cite{Crewther,Vecchia} yield
\begin{equation}
|d_N| \sim \theta \times O(10^{-2}\sim 10^{-3}) \,e\cdot{\rm fm},
\end{equation}
which leads to a bound $\theta \le O(10^{-10})$.
Hence $\theta$ must be very small or even may vanish in QCD.

Smallness of $\theta$ in the QCD sector, however,
is not protected in the presence of the
electroweak sector of the standard model, where the quark mass matrix,
arising from Yukawa couplings to the Higgs field,
may be written as
\begin{equation}
  \bar\psi^R_i(x) M_{ij}\psi^L_j(x) + \bar\psi^L_i(x)
M^{\dag}_{ij}\psi^R_j(x),
\end{equation}
where $\psi^L$ and $\psi^R$ represent left and right handed quark
fields
with flavor indices $i,j$.  Diagonalizing the mass matrix and
making it real, the parameter $\theta$ becomes
\begin{equation}
  \theta = \theta_{\rm QCD} + \arg\det M ,
\end{equation}
where $\theta_{\rm QCD}$ is the original $\theta$ parameter in QCD.
Therefore, $\theta_{\rm QCD}$ and $\arg\det M$ contributions 
have to cancel out to the precise degree that the stringent experimental upper 
bound on NEDM is satisfied.
In either of the two cases, it seems necessary to explain why Nature
chooses such a small value for $\theta$; this is the ``strong CP problem''.
One of the most attractive explanations proposed so far is the
Peccei-Quinn mechanism\cite{Peccei}.  Unfortunately,
the axion, a new particle predicted by this mechanism,
has not been experimentally observed so far.

Present theoretical estimations of NEDM vary 
in magnitude among 
different models such as current algebra\cite{Crewther},
chiral perturbation theory\cite{Vecchia,Aoki_Hatsuda,Cheng,Pich}, and
QCD sum rule\cite{Pospelov,Chan} (see also \cite{Pospelov:05}).
While these crude estimations of $|d_N|/\theta$ already convince 
the smallness of $\theta$, 
a theoretically reliable and accurate estimation for NEDM
will be required to determine the value of $\theta$,
if a non-zero value of  NEDM is observed in future experiments.
Lattice QCD calculations provide a first-principle method for this task.
Indeed more than fifteen years ago, the first attempt was made to estimate 
NEDM in a quenched lattice QCD simulation\cite{Aoki1}.
Reliable signal of NEDM could not be obtained at this time\cite{Aoki2}.
Since then, no lattice calculation of NEDM have been attempted until recently.
In the last year, new approach has been presented for this problem.
Ref.\cite{Shitani_form,Shintani_wE,Shintani_wE2} proposed a formulation to extract
the CP-odd electromagnetic form factor of nucleon from certain lattice 
correlation functions. NEDM can be extracted from this form factor in the zero 
momentum transfer limit. Applying this formulation in a quenched calculation
with domain-wall quarks, a non-zero value for the CP-odd form factor of nucleon
was obtained at one value of non-zero momentum.
Based on this formulation, the same form factor has been calculated on 
gauge configurations generated by $N_f=2$ dynamical domain-wall QCD
at several non-zero momenta\cite{RBC}. The value of NEDM after the zero 
momentum extrapolation, however, 
is consistent with zero within the large statistical error in this calculation.

The results mentioned above suggest that, while it is possible to 
obtain signals for the CP odd form factor at fixed and small value of momentum,
it is numerically difficult to 
carry out a statistically controlled extrapolation of the form factor to the zero momentum limit to
extract the value of EDM.
Therefore, in this paper, we investigate another method to calculate
the value of EDM directly without momentum extrapolation.
In this method, introducing a constant uniform electric field $\vec E$,
we measure the energy difference between spin-up and spin-down
components of the nucleon in the presence of the $\theta$ term~\cite{Aoki1}.
If the electric field is small enough, the leading contribution to the energy
difference is given by $d_N\vec S\cdot \vec{E}$ with neutron spin $\vec S$ and 
electric field $\vec E$. Therefore EDM can be directly
extracted without momentum extrapolation.
The most difficult part of this calculation is to reweight
the nucleon propagator on a given gauge configuration
with the factor $e^{i\theta Q}$, where $Q$ is the topological charge of the
configuration. We may control this reweighting by taking a small value of
$\theta$.
Another difficulty is that 
our electric field breaks periodicity in the time 
direction, generating large field at the time boundary. We should investigate
influences of the large electric field at the boundary to EDM signals.

We check 
the ability of this method in the quenched approximation at a heavy quark mass. 
We employ two fermion formulations,
domain-wall fermion having chiral symmetry and clover fermion with explicitly broken 
chiral symmetry, in order to investigate possible dependence of EDM 
signals on the aspect of chiral symmetry of fermion formulations.
Our study have revealed that the quality of EDM signals is not very 
sensitive to fermion formulations.
Therefore we have employed the clover fermion, which requires much less 
computational cost than the domain-wall fermion,
to study various systematics of EDM such as the volume dependence, the 
boundary effect and the quark mass dependence within the quenched 
approximation.

This paper is organized as following. 
In sec.~\ref{sec:NEDM_E_field} we explain the definition of EDM 
and our method to extract EDM from nucleon propagators.
Simulation details of our lattice calculation are summarized in
sec.~\ref{sec:simulation}.
In sec.~\ref{sec:result_Efield} we show numerical results with both 
domain-wall and clover fermion at heavy quark mass on a $16^3\times 32$ 
lattice. We then investigate the finite size effect and the boundary effect 
on a $24^3\times 32$ lattice with the clover fermion. 
In \ref{sec:quark_mass_dep} we systematically study the quark mass dependence 
of EDM using the larger lattice with the clover fermion. 
A summary and discussion is given in the last section \ref{sec:summary}.


\section{EDM with electric field}\label{sec:NEDM_E_field}
In our previous work\cite{Shitani_form},
we defined NEDM from the CP-odd electromagnetic form factor, $F_3$,
in the zero momentum transfer limit. 
In the actual calculation, however, it is not so easy to change the momentum transfer,
since the momentum is quantized as $\displaystyle p=n\frac{\pi}{L}$ 
on a finite spatial length of $L$.  
In the case of large $p$ with $n=2,3,\cdots$ at small $L$, statistical errors are large,
while a smaller $p$ with $n=2,3,\cdots$, which has a better signal, requires a larger lattice size $L$.
In both cases, the calculation becomes more difficult for larger momentum at $n=2,3\cdots$ than for
the smallest momentum at $n=1$. In addition,
the correct distribution of the topological charge is essentially  important for 
the NEDM calculation. 
Since the width of the distribution of topological charge increases linearly with
the volume, larger volume calculations require more statistics than the small volume ones, 
contrary to other observables. 

The difficulties for the extrapolation to the zero momentum transfer limit mentioned above
are our motivations to consider a different method for the NEDM calculation with which
we can avoid the momentum extrapolation.
In this section we introduce our new approach for the lattice QCD calculation of NEDM. 

\subsection{Formulation}
In ref.~\cite{Aoki1}, NEDM is defined through the energy change of the neutron state 
in the presence of an external electric field, similarly to the magnetic moment defined 
from that in the magnetic field.
If a static and uniform electric field 
exists in a CP-violating system, the electric dipole moment(EDM) appears in the Hamiltonian as 
the interaction term between spin $\vec S$ of particle and electric field $\vec E$:
\begin{equation}\label{eq:EDM_hamil}
  \delta H_{\rm CP} = d_N(\theta){\vec{S}}\cdot\vec{E} + O((\vec E)^3)
\end{equation}
where $d_N(\theta)$ represents the EDM.
In order to extract EDM we consider the energy difference of nucleon states for
opposite spins in the external electric field:
\begin{equation}
  m^\theta_{\vec{s}}(\vec{E}) - m^\theta_{-\vec{s}}(\vec{E}) 
  = 2 d_N(\theta){\vec{S}}\cdot\vec{E}+ O((\vec E)^3), 
\label{eq:EnergyDiff}
\end{equation}
where $m^\theta_{\pm\vec{s}}(\vec{E})$ denotes the energy of nucleon
whose spin vector is $\pm\vec{S}$
in the presence of the electric field $\vec{E}$. 
Therefore we can extract $d_N(\theta)$ from the nucleon propagators for two different spin
states at zero momentum only, avoiding difficult calculations at non-zero momenta.

For small $\theta$ we can expand $d_N(\theta)$ as
\begin{equation}
  d_N(\theta) = d_N \theta + \mathcal O(\theta^3). 
  \label{eq:small_theta}
\end{equation}
We will check that higher order contributions at
$\mathcal O(\theta^3)$ are negligibly small. 
Hereafter we represent $d_N$ as the leading order of EDM.

\subsection{Methodology on the lattice}
A static and uniform electric field is represented by the spatial gauge potential as
\begin{equation}
  A_i ( x) = E_i^{\rm Euclid} t
\end{equation}
where $E_i^{\rm Euclid}$ is the constant electric field in Euclidean space.
A non-zero NEDM could be detected from
the oscillating behavior of the neutron propagator. Since NEDM is expected to be small,
it is numerically very difficult to measure such a small oscillation.
On the other hand, if we employ a static and uniform electric field in Minkovski  space as
\begin{equation}
  A_i (x) = -i E_i^{\rm Minkov} t,
\end{equation}
the oscillation turns into an exponential behavior, which is easier to measure.
Therefore we  introduce a static and uniform electric field in Minkovski
space  as an external field into lattice QCD, by replacing the spatial link variables 
as 
\begin{eqnarray}
  U_i(x) \longrightarrow \widetilde U_i(x;E^{\rm Minkov}_i) 
         &=& e^{q_eE^{\rm Minkov}_it}U_i(x),\nonumber\\
  U^{-1}_i(x) \longrightarrow \widetilde U_i^{-1}(x;E^{\rm Minkov}_i) 
         &=& e^{-q_eE^{\rm Minkov}_it}U^{-1}_i(x),
\label{eq:inc_E}
\end{eqnarray}
where $q_e$ denotes the quark charge, $2/3$ for up quark and $-1/3$ for down quark.
Hereafter
we suppress the superscript of the constant electric field $E_i$  in Minkovski 
space for simplicity.

An obvious problem here is that the Minkovski electric field $E$ breaks 
the periodic boundary condition in the temporal direction:
\begin{eqnarray}
  U_i(t+T,\vec x) &=& U_i(t,\vec x),\\ 
  \widetilde U_i(t+T,\vec x;E_i) &=& e^{q_eE_iT}\widetilde U_i(t,\vec x;E_i)
  \ne \widetilde U_i(t,\vec x;E_i)
\end{eqnarray}
where $T$ is the size of the temporal direction.
This generates an effective electric field, defined by $ E_i (t) = \frac{A_i(t+1)-A_i(t-1)}{2}$, as
\begin{equation}
  E_i (t) = \left\{
\begin{array}{ll}
   E_i  & t=2,3,\cdots, T-1 \\
  -E_i\frac{T-2}{2} & t=1, T \\
\end{array}\right. .
\end{equation}
Therefore the electric field is no more constant near the boundary between $t=1$ and $t=T$.
In order to avoid the effect of this non-uniform electric field to the EDM signal,
we have to take $E_i$ as small as possible.
In any case
a small value of $E_i$ is necessary to neglect $O((\vec{E})^3)$ terms in (\ref{eq:EDM_hamil}).
\footnote{The electric field in Euclidean space smaller than $E_i = 2\pi/T$
also breaks the periodic boundary condition.}

In our calculation
gauge configurations are generated by the usual lattice QCD action
without $E$ and $\theta$.
After inserting the electric field to gauge configurations 
we calculate quark propagators for flavor $u$ and $d$ separately,
in addition to the normal one with $E_i = 0$, which is used to remove a fake signal at $E_i=0$
caused by statistical fluctuations. 
The total number of solvers for quark propagators is three for each configuration.
From quark propagators we construct the nucleon propagator with the $\theta$ term as
\begin{equation}
    \langle N_{\alpha} \bar N_{\beta} \rangle_{\theta}(\vec E,t) 
  = \langle N_{\alpha}(t) \bar N_{\alpha}(0) e^{i\theta Q}\rangle_{\vec E}
\end{equation}
where $\langle {\cal O}\rangle_{\vec E}$ represents the vacuum expectation value of
${\cal O}$ with $\vec E$ but without the $\theta$ term. 
Here we use the re-weighting method with the complex weight factor $e^{i\theta Q}$.
In order to obtain good signals,
a large overlap of gauge ensembles between $\theta=0$ and $\theta\not= 0$
as well as the correct distribution of the topological charge are required.
Taking a small value of $\theta$ as long as we get a signal
helps for the large overlap, while
we have to simply accumulate an enough number of configurations
for the correct distribution of the topological charge.

In the presence of the uniform and static electric field, 
the upper components of the nucleon propagator at zero spatial momentum 
take the following form for $\alpha,\,\beta=1,2$\cite{Aoki3}:
\begin{eqnarray}
  \langle N_{\alpha} \bar N_{\beta} \rangle_{\theta}(\vec E,t) 
 &= & Z'_N{}^{\theta}(E^2\equiv\vec E\cdot \vec E)
    \Big[\left(1 + A_N(\theta, E^2)\vec\sigma\cdot\vec E\right)\nonumber\\
 &\times& \exp\big(-m_N^\theta(E^2)t 
  - \frac{d_N(\theta,E^2)}{2}\vec{\sigma}\cdot\vec{E}t\big)
    \Big]_{\alpha\beta} + \cdots ,
\label{eq:NN_with_E}
\end{eqnarray}
where  the EDM $d_N(\theta,E^2)$ and the spin-dependent amplitude
$A_N(\theta, E^2)$ are odd in $\theta$, while
the spin-independent energy\footnote{
The energy of the proton increases as $t$ increase since the charged particle
is accelerated in the uniform electric field.  This effect is canceled in the ratio,
which will be used to extract the signal of EDM.} $m^\theta(E^2)$
and an overall amplitude $Z'_N{}^{\theta}(E^2)$ are even in  $\theta$.  
Here dots denote contributions from excited states.

To extract EDM
we construct the ratio of nucleon propagators between different spinor components. 
For $\vec{E}=(0,0,E)$ we consider the following ratio:
\begin{equation}\label{eq:ratio_naive}
  R^{\rm naive}_3(E,t;\theta) = \frac{\langle N_1 \bar N_1 \rangle_{\theta}((0,0,E),t)}
                                     {\langle N_2 \bar N_2 \rangle_{\theta}((0,0,E),t)}
  = \frac{1+A_N(\theta,E^2) E}{1-A_N(\theta,E^2) E}
    \exp[ -d_N \theta Et + \mathcal O(\theta^3 E,\theta E^3) ],
\end{equation}
where we use eq.~(\ref{eq:NN_with_E}) for the second equality.
Similarly for  $\vec{E}=(E,0,0)$ and $(0,E,0)$, we obtain
\begin{eqnarray}
  R^{\rm naive}_1(E,t;\theta) &=& 
  \frac{\langle N_1\bar N_1\rangle_{\theta} + \langle N_1\bar N_2\rangle_{\theta}
      + \langle N_2\bar N_1\rangle_{\theta} + \langle N_2\bar N_2\rangle_{\theta}}
       {\langle N_1\bar N_1\rangle_{\theta} - \langle N_1\bar N_2\rangle_{\theta}
      - \langle N_2\bar N_1\rangle_{\theta} + \langle N_2\bar N_2\rangle_{\theta}}((E,0,0),t)
  \label{eq:dn_x}\\
  &=&  \frac{1+A_N(\theta,E^2) E}{1-A_N(\theta,E^2) E}\exp[ -d_N \theta Et  
   + \mathcal O(\theta^3 E, \theta E^3) ],\nonumber\\
  R^{\rm naive}_2(E,t;\theta) &=& 
  \frac{\langle N_1\bar N_1\rangle_{\theta} + i\langle N_1\bar N_2\rangle_{\theta}
     - i\langle N_2\bar N_1\rangle_{\theta} +  \langle N_2\bar N_2\rangle_{\theta}}
       {\langle N_1\bar N_1\rangle_{\theta} - i\langle N_1\bar N_2\rangle_{\theta}
     + i\langle N_2\bar N_1\rangle_{\theta} +  \langle N_2\bar N_2\rangle_{\theta}}((0,E,0),t)
  \label{eq:dn_y}\\
  &=& \frac{1+A_N(\theta,E^2) E}{1-A_N(\theta,E^2) E}\exp[ -d_N \theta Et  
   + \mathcal O(\theta^3E, \theta E^3) ] .\nonumber
\end{eqnarray}
We can average over the ratio in three directions to increase statistics, if necessary.

In order to remove the spurious contribution $m^\theta_{\vec{s}}(0)-m^\theta_{-\vec{s}}(0)$,
which must vanish for infinite statistics, we consider a double ratio defined by
\begin{eqnarray}
  R_i(E,t;\theta) 
  &=& \left[\frac{R^{\rm naive}_i(E,t;\theta)}{R^{\rm naive}_i(0,t;\theta)}\right],\\
  \ln\left[\frac{R_i(E,t;\theta)}{R_i(E,t+1;\theta)}\right] 
  &=& [m^\theta_{\vec s}(E_i)-m^\theta_{\vec s}(0)] 
    - [m^\theta_{-\vec s}(E_i)-m^\theta_{-\vec s}(0)]\\
  &=& d_N\theta E + \mathcal O(\theta^3 E, \theta E^3).
\end{eqnarray}
We can improve the EDM signal further, removing the contribution at $\theta = 0$,
which also vanish for infinite statistics, by a triple ratio as
\begin{equation}\label{eq:ratio_wo_theta0}
  R_i^{({\rm w/o\,\theta=0})}(E,t;\theta)
    = \frac{R_i(E,t;\theta)}{R_i(E,t;\theta=0)}
  \simeq \frac{1+\theta A_N^1(E^2) E}{1-\theta A_N^1(E^2) E}\exp[d_N\theta Et],
\end{equation}
where we used an expansion
$A_N(\theta,E^2) = \theta A_N^1(E^2) + \mathcal O(\theta^3)$, 
and we finally subtract the spurious contribution even  in $E$ by a quadruple ratio as
\begin{eqnarray}
  R_i^{\rm corr}(E,t;\theta) 
&=& \frac{R^{({\rm w/o\,\theta=0})}_i(E,t;\theta)}{R^{({\rm w/o\,\theta=0})}_i(-E,t;\theta)}
    \nonumber\\
&=& \frac{R_i^{\rm naive}(E,t;\theta)}{R_i^{\rm naive}(-E,t;\theta)}
    \frac{R_i^{\rm naive}(-E,t;\theta=0)}{R_i^{\rm naive}(E,t;\theta=0)}\nonumber\\
&\simeq& \left(\frac{1+\theta A_N^1(E^2)E}{1-\theta A_N^1(E^2)E}\right)^2\exp[2d_N\theta E t],
\label{eq:ratio_wo_theta0_E}
\end{eqnarray}
where the second equality tell us that this is indeed a triple ratio 
since $\vec E=0$ contributions are canceled identically. 
We finally extract EDM from the exponential fit to $R_i^{\rm corr}(E,t;\theta)$ over
some time range, determined by the behavior of  the effective EDM:
\begin{equation}
  2d_N\theta E = \ln\left[\frac{R^{\rm corr}_i(E,t;\theta)}{R^{\rm corr}_i(E,t+1;\theta)}\right],
  \quad i=1,2,3.
\end{equation}


\section{Simulation details}
\label{sec:simulation}
\subsection{Simulation parameters} 
In our study we employ gauge configurations generated by the RG improved gauge action
at $\beta=2.6$ in the quenched approximation, which corresponds to $a^{-1} = 1.902(50)$ GeV
from the string tension  $\sigma$ assuming $\sigma = (440\,\textrm{MeV}^2)$\cite{Okamoto}.

For the quark action, we employ the domain-wall fermion on a $16^3\times 32$ lattice 
with the fifth length $N_s=16$ and the domain-wall height $M=1.8$.
These parameters are identical to those in the previous EDM form factor calculation.
We however take a heavier quark mass, $m_f=0.12$, which corresponds to $m_Na=1.113(2)$ and 
$m_{PS}/m_{V}\simeq 0.88$, than the one in the previous calculation, in order to reduce the
computational cost, since our main motivation in this calculation is to see whether the EDM signal
can be obtained by this method.  As shown in the next section we have indeed obtained 
the EDM signal after accumulating 1000 configurations at this heavier quark mass. 

We also investigate whether the EDM signal can be obtained by this method with the clover fermion.
The EDM calculation with this fermion has the advantage that the computational cost is
roughly $N_s$ times smaller than the cost of the domain-wall fermion 
so that systematic studies such as
volume or quark mass dependences can be performed more easily.
Moreover we can employ the $N_f=2$ and $2+1$ flavor dynamical configurations 
already generated with the clover quark action at several sea quark masses 
and lattice spacings\cite{cppacs, cppacs-jlqcd, pacscs} in future studies. 
We calculate the EDM on the same $16^3\times 32$ configurations, using
the clover fermion with $c_{SW}=1.340$,
the tadpole improved value of the clover coefficient determined  from
\begin{equation}\label{eq:c_sw}
  c_{SW} = \Big[\sum_{x,\mu<\nu}P_{\mu\nu}(x)\Big]^{-3/4} 
         = (1-0.8412\beta^{-1})^{-1/4} .
\end{equation}
In order to obtain a similar nucleon mass, we use 
the hopping parameter $\kappa=0.1320$, corresponding to $m_Na=1.020(2)$
and $m_{PS}/m_{V}\simeq 0.85$.

Since, as will be shown later, the EDM signal can be successfully obtained with the clover fermions,
we investigate the volume dependence of the EDM signal using a $24^3\times 32$ lattice.
Furthermore the quark mass dependence of the EDM is calculated with this fermion
on this larger volume. 

\begin{figure}[h]
\begin{center}
\vskip 10mm
\includegraphics[width=70mm, angle=0] {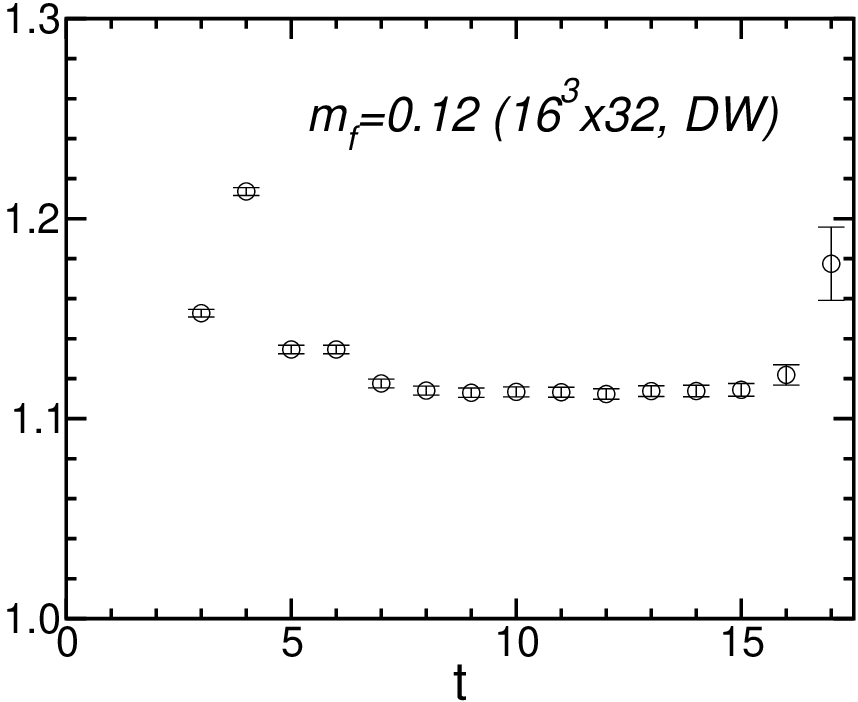}
\hspace{0.5cm}
\includegraphics[width=70mm, angle=0] {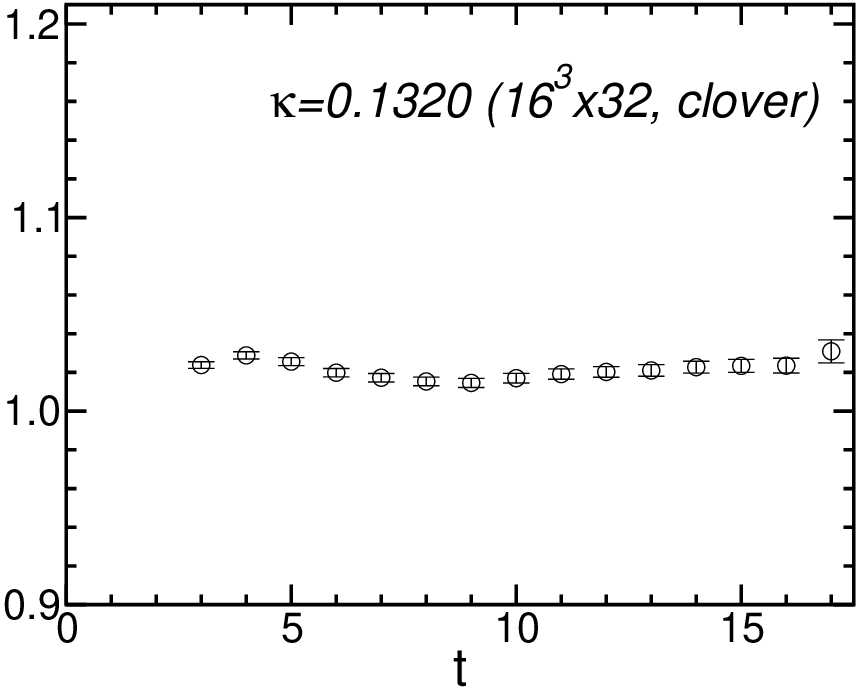}
\vskip 1cm
\includegraphics[width=100mm, angle=0] {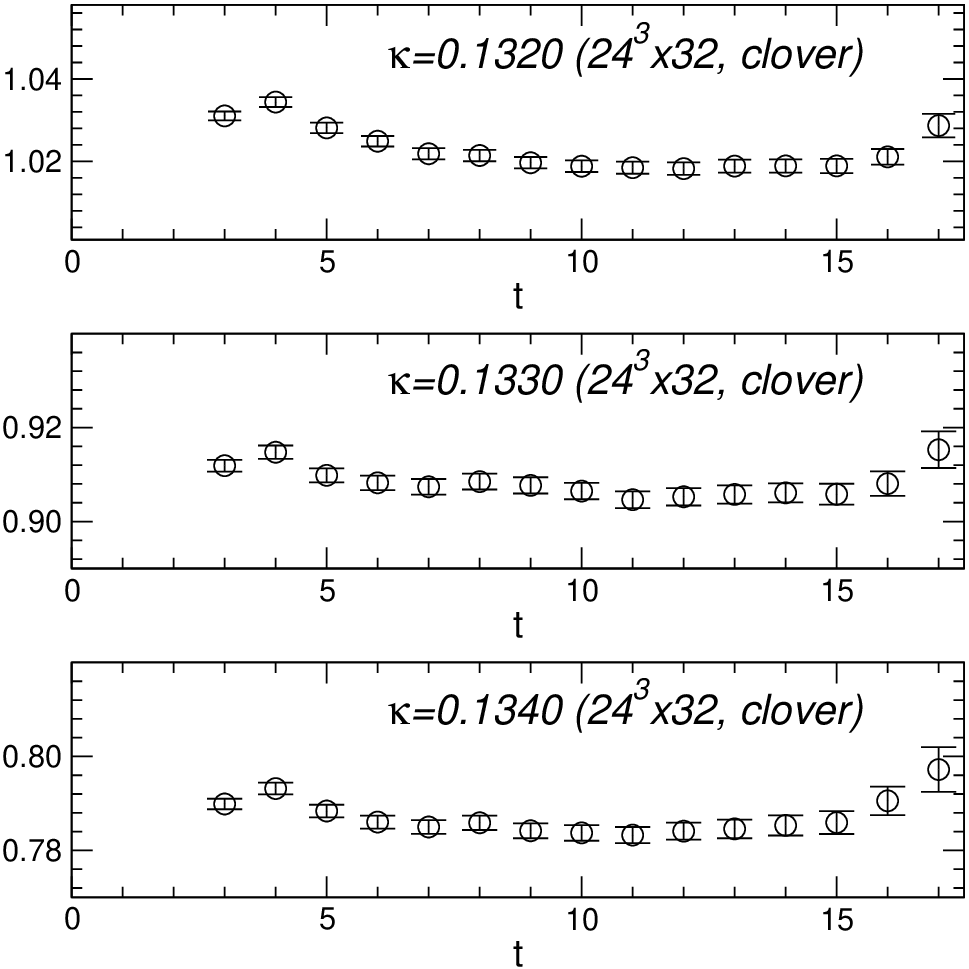}
\caption{The effective mass plot for nucleon with domain-wall fermion at $m_f=0.12$ (top-left), 
clover fermion at $\kappa=0.1320$ (top-right) on a $16^3\times 32$ lattice and clover fermion 
at various quark masses (bottom) on a $24^3\times 32$.}
\label{fig:Nmass}
\end{center}
\end{figure}

For the calculation of quark propagators we employ the smeared source
of the form that $f(r)=Ae^{-Br}$ where $ r=\vert\vec{x} -\vec x_{\rm src}\vert$ with the source point
$\vec x_{\rm src} =(8,8,8)$ on $16^3$ and $(12,12,12)$ on $24^3$ spatial lattice,
after the Coulomb gauge fixing is applied to gauge configurations.
We mainly take $t_{\rm src} = 1$ as the time slice of the smeared source. 
In order to check the effect of the non-uniform electric field near $t=1$ and $T$, we also
calculate the EDM with $t_{\rm src}=8$, using the clover fermion on a $24^3\times 32$ lattice.
Effective mass plots of nucleon in various cases are given in Fig.~\ref{fig:Nmass}.
We observe the plateau at $ t \ge 7$ for the domain-wall fermion and 
the clover fermion at heaviest quark mass, while plateau 
appears at $ t\ge 6$ for the clover fermion at lighter quark masses.

In our calculation we mainly take $\vec{E} = (0,0,E)$ with $E=\pm 0.004$. 
As exceptions,
$E=\pm 0.002$ is employed on a $16^3\times 32$ lattice with the domain-wall fermion
to investigate the $E$ dependence of the EDM signal, and
 $(E, 0, 0)$ and $(0,E,0)$ are used on a $24^3\times 32$ lattice with the clover fermion
 at heaviest quark mass to check the consistency and to increase statistics. 
Although we can easily change the value of $\theta$ by reweighting, we fix $\theta = 0.1$ in
our calculation, except $\theta = 0.05$ and $0.2$ on a $16^3\times 32$ lattice 
with the domain-wall fermion to investigate the $\theta$ dependence of the EDM signal.

Parameters of fermion actions in various cases are summarized in Table~\ref{tab:lat_param}.

\subsection{Topological charge}
The topological charge using the $\mathcal O(a^2)$ improved definition\cite{impQ}
is measured on each configuration after 20 cooling steps. 

\begin{figure}[h]
\begin{center}
\vskip 10mm
\includegraphics[width=90mm, angle=0] {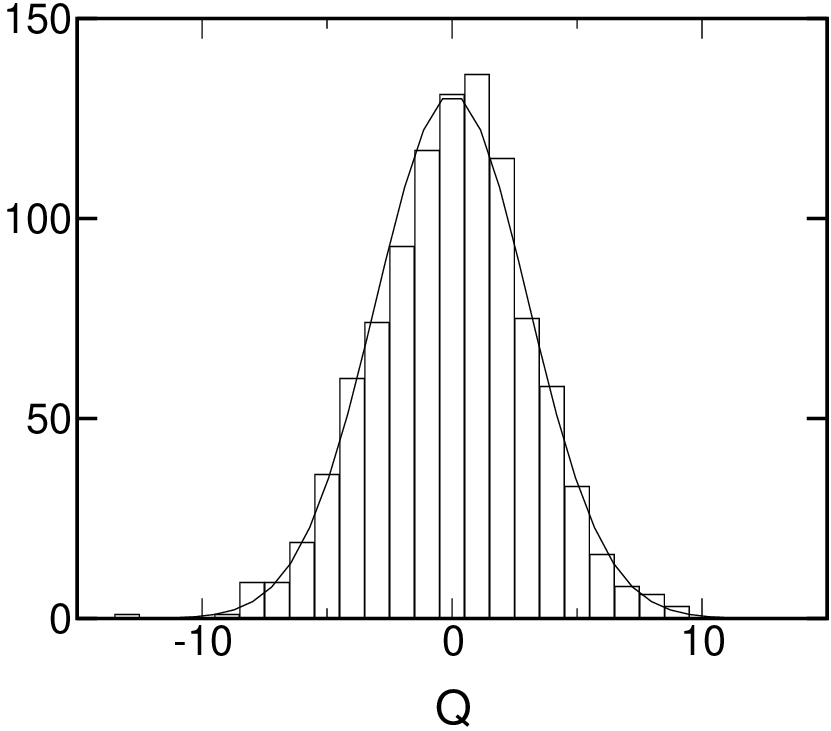}
\vskip 10mm
\includegraphics[width=150mm, angle=0] {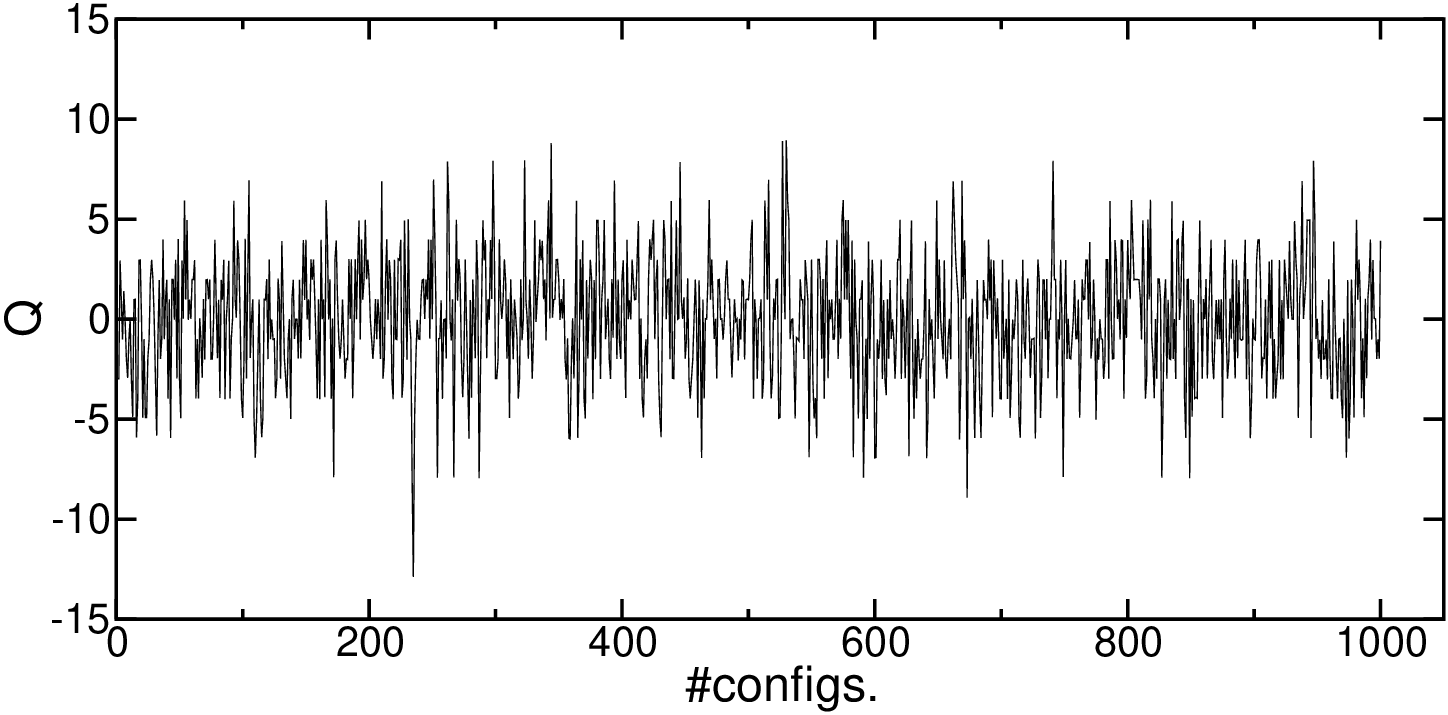}
\caption{(top) Histogram of topological charge improved by $\mathcal{O}(a^2)$ 
after 20 cooling steps. The solid line denotes the expected gaussian distribution
from $\sigma = \sqrt{\langle Q^2\rangle - \langle Q\rangle^2}$.
(bottom) The topological charge in each configuration.}
\label{fig:hist_Q}
\end{center}
\end{figure}

On  a $16^3\times 32$ lattice we accumulate 1000 configurations.
In Fig.~\ref{fig:hist_Q} we present the histogram of the topological charge,
which is consistent with gaussian distribution.
The symmetry of the distribution is measured by the average of $Q$, which is consistent with 
zero within error:  $\langle Q \rangle = -0.002(97)$. 
If the gaussian distribution is assumed, its width $\sigma$
is given by $\langle Q^2\rangle = 9.37(44)$.
On this lattice size 1000 configurations seem enough to give a reasonable distribution 
of the topological charge.

\begin{figure}[h]
\begin{center}
\vskip 10mm 
\includegraphics[width=100mm, angle=0] {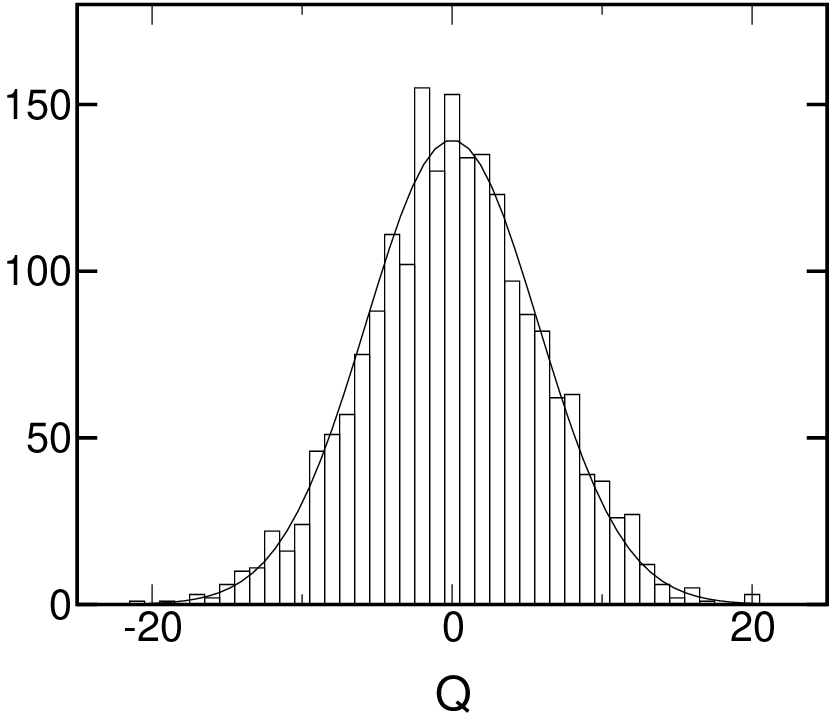}
\vskip 10mm
\includegraphics[width=150mm, angle=0] {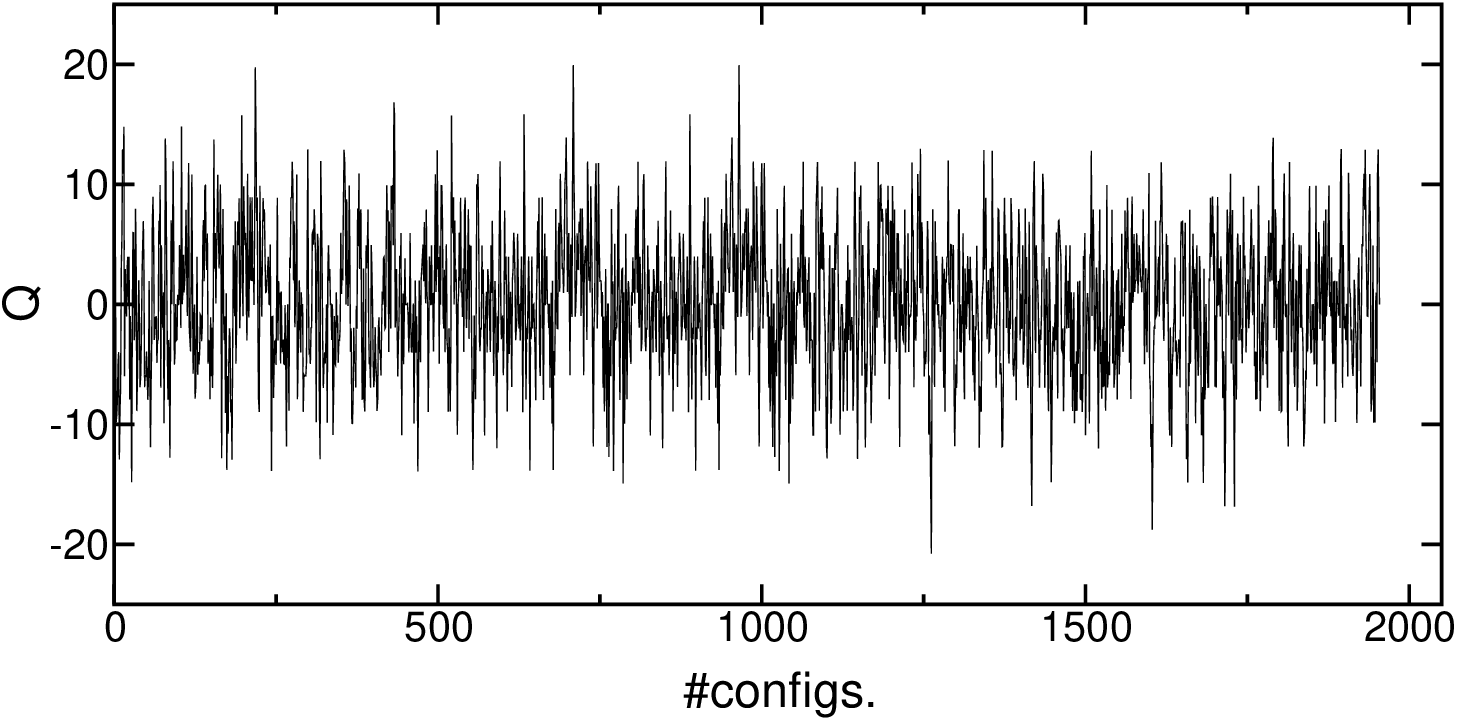}
\caption{(top) Histogram of the topological charge in $24^3\times 32$ lattice
and (bottom) the topological charge in each configuration 
as shown in Fig.~\ref{fig:hist_Q}.}
\label{hist_Ns32}
\end{center}
\end{figure}

On a larger volume of a $24^3\times 32$ lattice, we accumulate nearly 2000 configurations 
since $\langle Q^2\rangle $, thus the width of the distribution of $Q$, increases linearly in volume.
In Fig.~\ref{hist_Ns32}, we show the histogram of $Q$,
which looks reasonable,
namely sufficiently symmetric and close to gaussian. 
We find $\langle Q\rangle = 0.15(13)$ and $\langle Q^2\rangle = 33.6(1.1)$.

\section{EDM signal and Systematics }\label{sec:result_Efield}
In this section, we show numerical results for nucleon EDM signals with the external 
electric field method. We investigate several systematics of the EDM signal such as
dependences on
the fermion action, the volume, $E$,  $\theta$, $t_{\rm src}$ and the direction
of $\vec E$. 
  
\subsection{Comparison between domain-wall and clover fermions}\label{sec:EDM_clvDW}
\begin{figure}[h]
\begin{center}
\vskip 10mm
\includegraphics[width=100mm, angle=0] {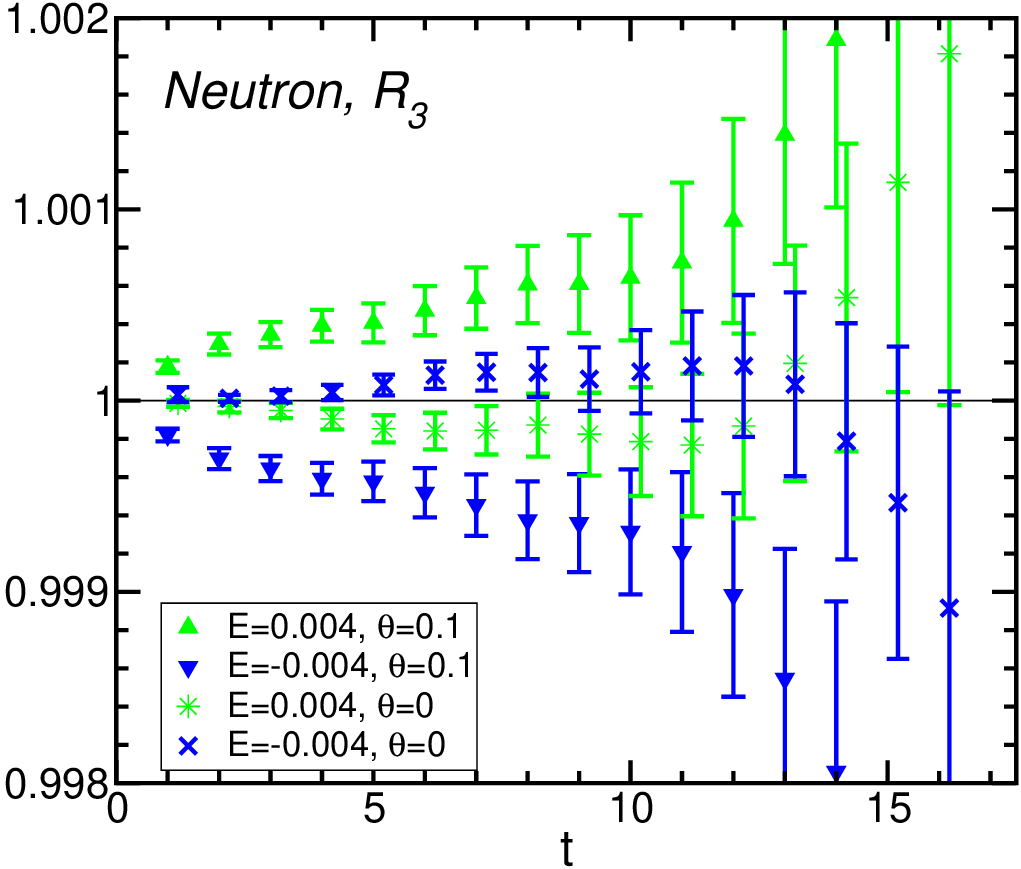}
\vskip 10mm
\includegraphics[width=100mm, angle=0] {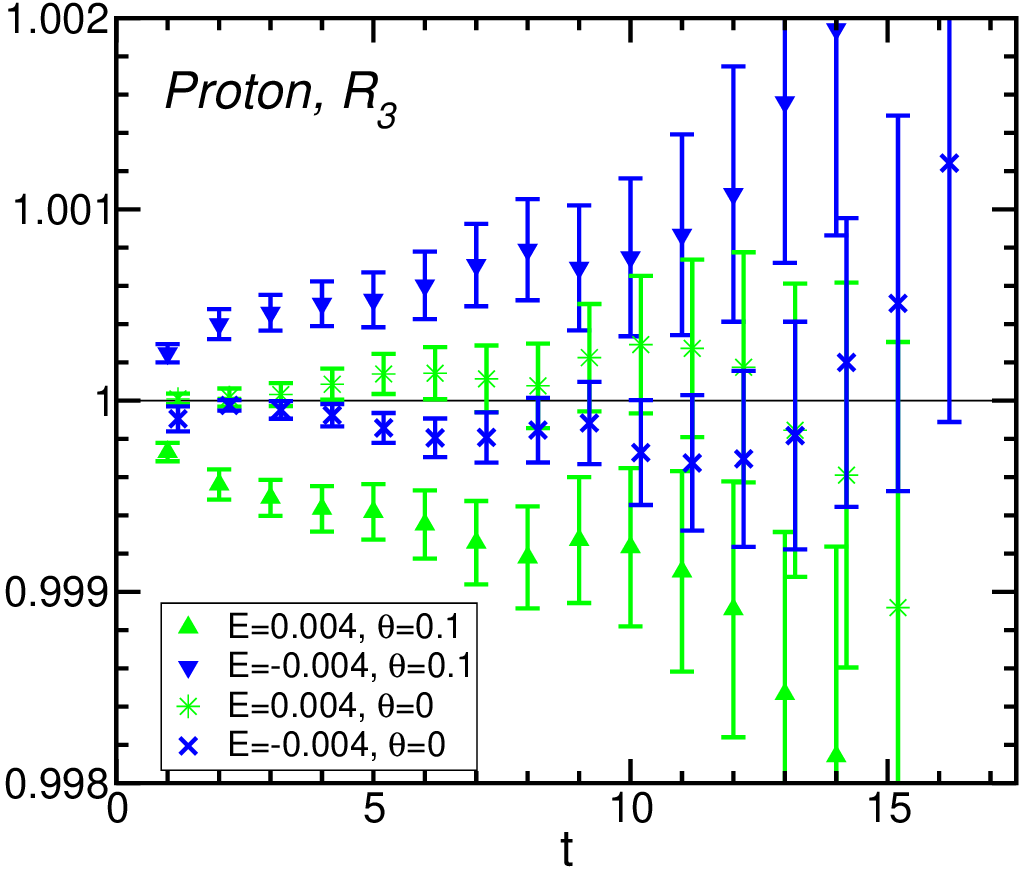}
\caption{The time behavior of $R_3(E,t;\theta)$ in $E=\pm 0.004,\,\theta=0.1$ 
and $E=0.004,\,\theta=0$ with domain-wall fermion. 
(Top) neutron case, (bottom) proton case.}
\label{fig:Edn_ratio_DW}
\end{center}
\end{figure}

\begin{figure}[h]
\begin{center}
\vskip 10mm
\includegraphics[width=100mm, angle=0] {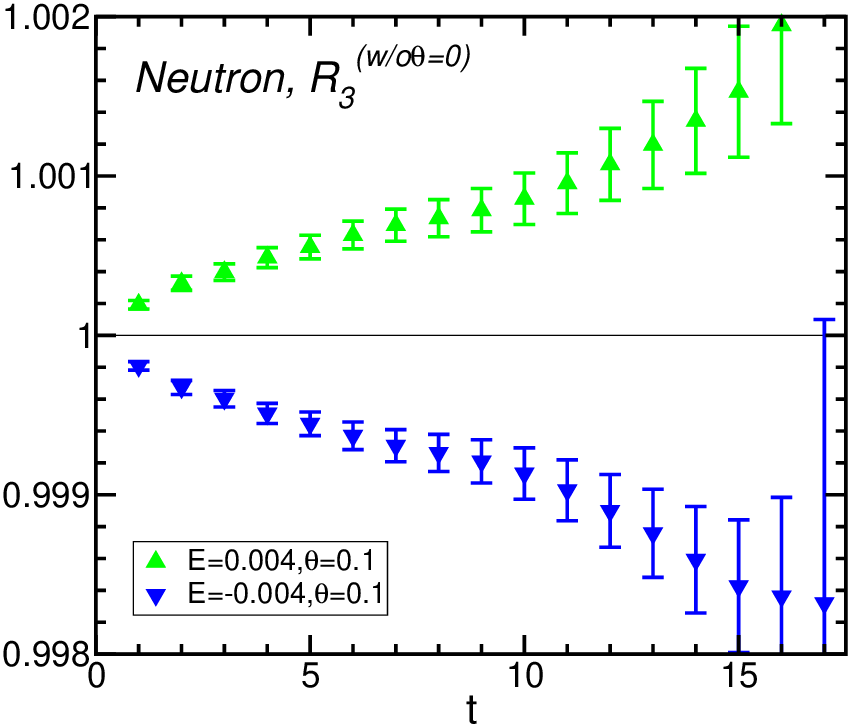}
\vskip 10mm
\includegraphics[width=100mm, angle=0] {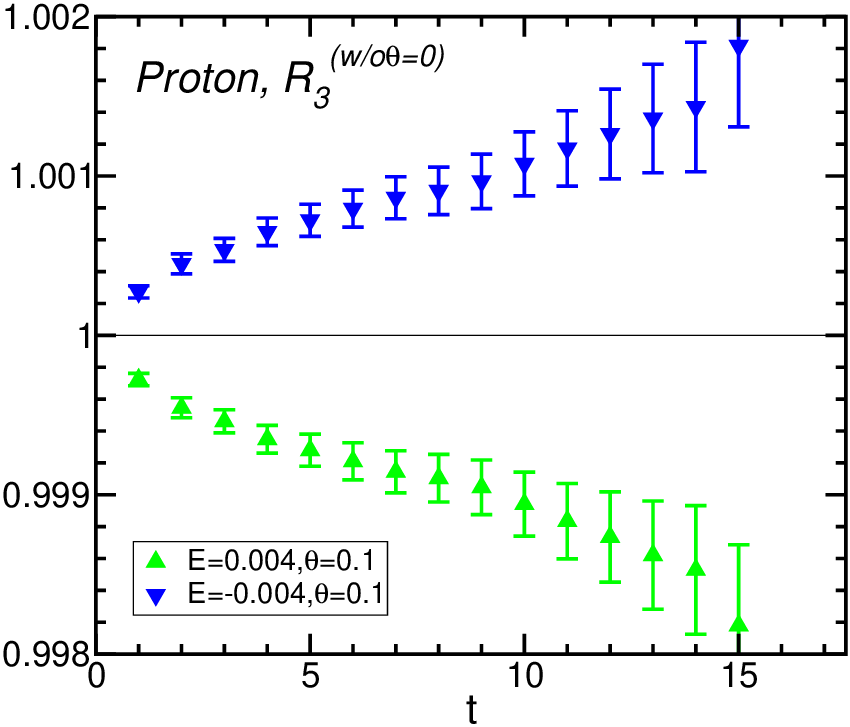}
\caption{The time behavior of $R^{(\rm w/o\theta=0)}_3(E,t;\theta)$ 
in $E=\pm 0.004,\,\theta=0.1$ with domain-wall fermion. 
(Top) neutron case, (bottom) proton case.}
\label{fig:Edn_ratio_DW_corr}
\end{center}
\end{figure}

We first consider the case of the domain-wall fermion on a $16^3\times 24$ lattice.
In Fig.~\ref{fig:Edn_ratio_DW} we plot the double ratio $R_3(E,t;\theta)$ as a function of $t$
at  $(E,\theta)=(\pm4.0\times 10^{-3},0.1)$ and $(E,\theta)=(\pm 4.0\times 10^{-3},\theta=0)$,
for both neutron and proton.
The star symbols in Fig.~\ref{fig:Edn_ratio_DW}, representing 
the time dependence of $R_3(\pm E,t;\theta=0)$, are consistent with unity within errors 
at both $\pm E$. 
This confirms the expected behavior that the exponential part of $R_3(E,t;\theta)$ vanishes
at $\theta = 0$.
For non-zero $\theta$, on the other hand,
deviations of  $R_3(E,t;\theta)$ from unity show up beyond errors and they
increases as $t$ increases.
Moreover the sign of deviations depends on the sign of $E$.
All these behaviors of $R_3$ are consistent with the fact that non-zero value of EDM exists.
In Fig.~\ref{fig:Edn_ratio_DW_corr} we plot time dependence of $R_3^{(\rm w/o \theta=0)}(E,t;\theta)$,
defined in eq.~(\ref{eq:ratio_wo_theta0}),
for which contributions at $\theta =0$ due to finite statistics are removed. 
The $E$ dependence of signals become more visible after the removal of $\theta=0$ contributions. 
In addition it is noted that the EDM signal of proton has an opposite sign to that of neutron. 

\begin{figure}[h]
\begin{center}
\vskip 10mm
\includegraphics[width=100mm, angle=0] {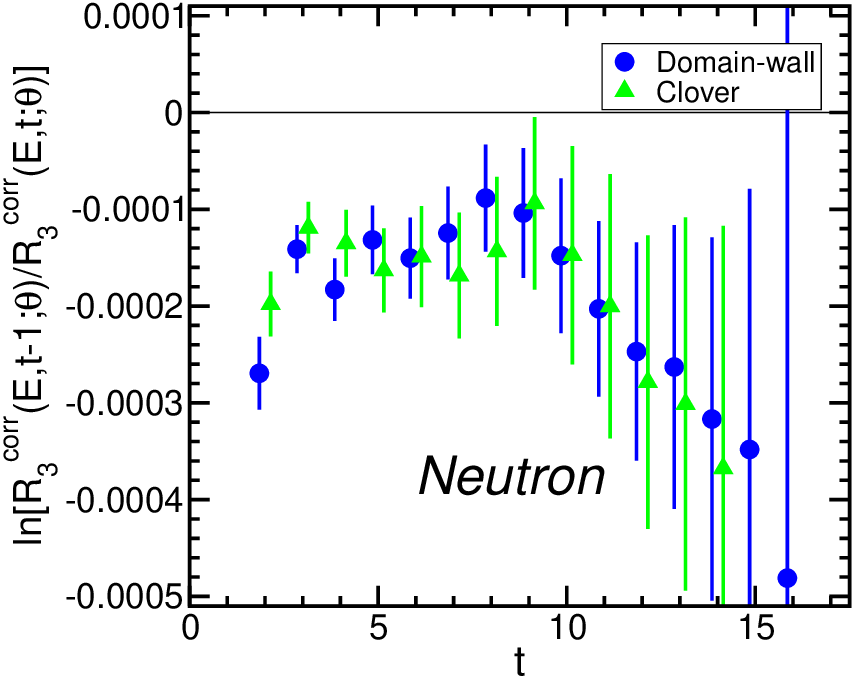}
\vskip 10mm
\includegraphics[width=100mm, angle=0] {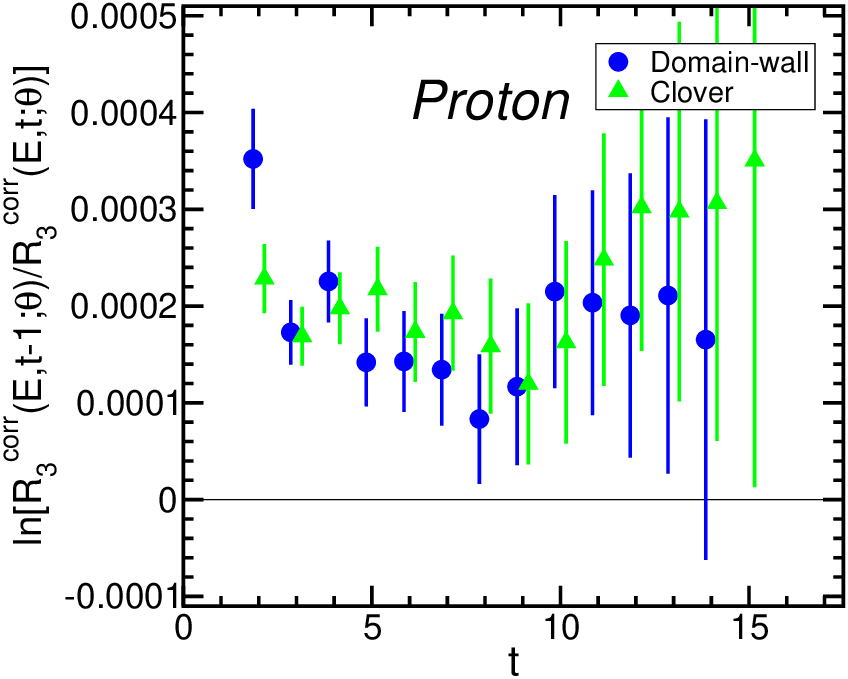}
\caption{The effective mass plot of $R^{\rm corr}_3(E,t;\theta)$ 
in $E=0.004,\,\theta=0.1$ with domain-wall and clover fermions in $16^3\times 32$ lattice. 
(top) neutron case, (bottom) proton case. }
\label{fig:Edn_effratio_clvDW}
\end{center}
\end{figure}

Applying the same analysis as above to the case of the clover fermion on a $16^3\times 32$
lattice, we obtain a similar behavior for $R_3$ and $R_3^{(\rm w/o \theta=0)}$.
Therefore we do not present them here.
Instead the effective mass of $R^{\rm corr}_3$, defined 
in eq.~(\ref{eq:ratio_wo_theta0_E}), is plotted as a function of $t$
in Fig.~\ref{fig:Edn_effratio_clvDW}, for both domain-wall and clover fermions.
It is interesting to see that the time dependences of the effective mass for the two fermions are
very similar.
Moreover, for both fermions, we observe plateau around $6\le t \le 12$,
whose values are non-zero beyond errors.
Clearly the EDM signal for proton has an opposite sign to that for neutron, as suggested
by the behavior of $R_3^{(\rm w/o \theta=0)}$.

Let us conclude this subsection.
Using the external electric field method, we obtain the EDM signal for both neutron and proton,
with both domain-wall and clover fermions.
This suggests that the chiral property of the fermion action does not
play a crucial role to obtain the EDM signal with this method.
Note however that the quark mass employed in this investigation is
rather heavy. Therefore there is a possibility that some qualitative 
difference between two fermion formulations may show up at lighter quark
mass where the chiral symmetry becomes important.
In the remaining of this paper, 
we mainly employ the clover fermion formulation.

\subsection{Volume dependence }\label{sec:size_eff}
We investigate the volume dependence of the EDM signal on a $24^3\times 32$ lattice
with the clover fermion at the heaviest quark mass.
Here the physical spatial volume is increased to $2.4^3\,\textrm{fm}^3$ from
$1.6^3\,\textrm{fm}^3$.
Our main concern is whether the nonzero value of the EDM signal obtained in the previous
subsection persists as the volume increases.

\begin{figure}[h]
\begin{center}
\vskip 10mm
\includegraphics[width=100mm, angle=0] {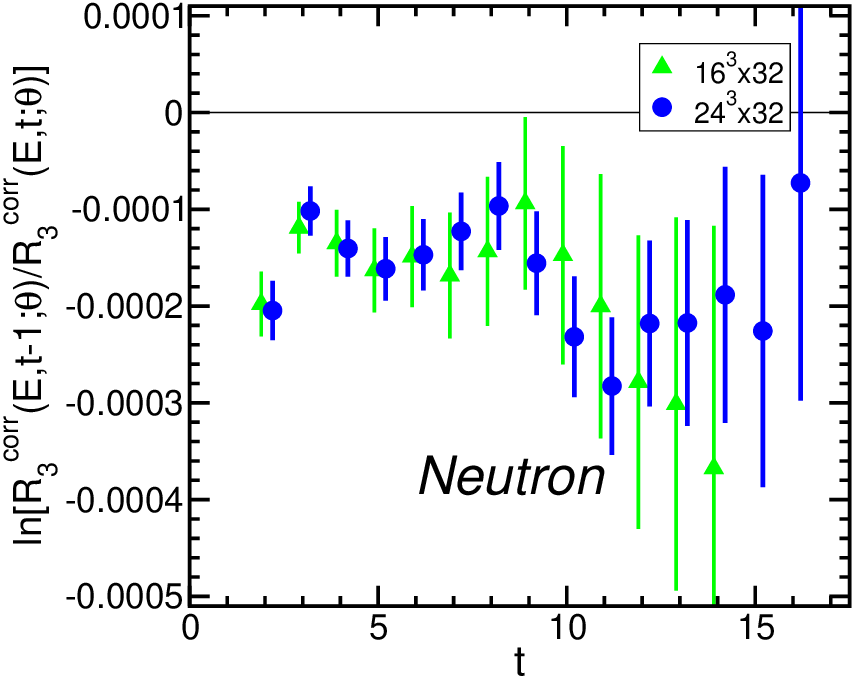}
\vskip 10mm
\includegraphics[width=100mm, angle=0] {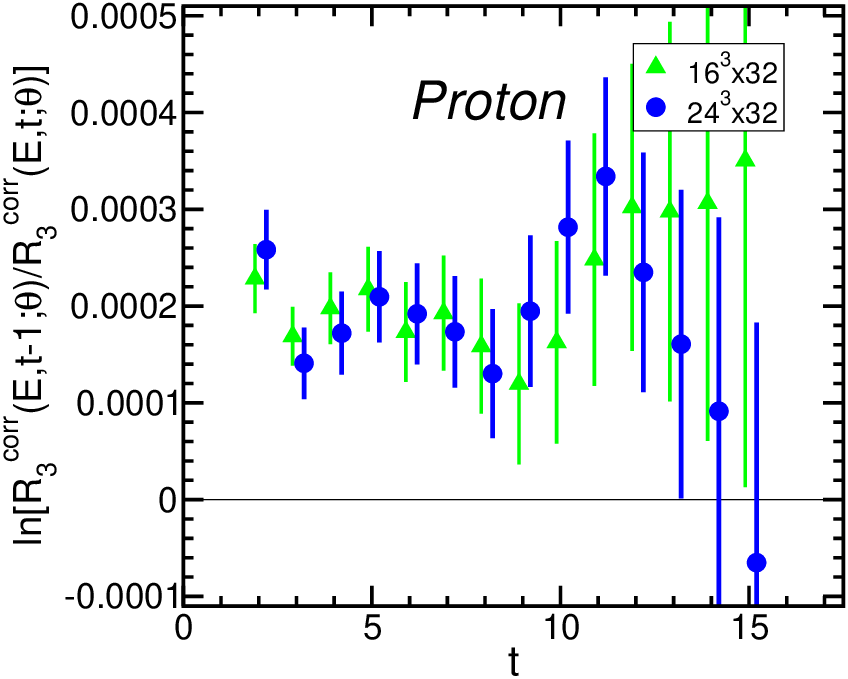}
\caption{The same figure as Fig.~\ref{fig:Edn_effratio_clvDW} for clover fermion in 
$16^3\times 32$ and $24^3\times 32$ lattice. (top) neutron case, (bottom) proton case. }
\label{fig:Edn_effratio_clv_Ns24}
\end{center}
\end{figure}

In Fig.~\ref{fig:Edn_effratio_clv_Ns24} we compare the effective mass plot of 
$R^{\rm corr}_3(E,t;\theta)$ at $\theta=0.1,\,E=0.004$
in the larger volume with that in the smaller volume.
It is clear that the EDM signal remains non-zero in the larger volume. 
Results in both volumes are consistent with each other within large errors.
We can conclude that the EDM signal obtained with this method does not vanish 
in both volumes.

\subsection{Boundary effect of the electric field }\label{sec:bc_eff}
The electric field in our method breaks periodicity in the
time direction, leading to a large non-uniformity near the boundary between $t=1$ and $t=T$. 
Since we put a source at $t=1$,
the EDM signal may be affected by the non-uniform electric field.
In order to investigate how the EDM signal is affected by this boundary effect,
we repeat the EDM calculation on a $24^3\times 32$ with the clover fermion
at the heaviest quark mass, moving the source point to the different time slice
but keeping other conditions fixed.

In the previous calculation at $t_{\rm src}=1$, we observed that the plateau seems to exist
at $ t \ge 8 $. 
Since this indicates that the effect of boundary may be small at $t=8$,
we take a new source point at $t_{\rm src}=8$. 
If we need a minimum plateau length of 5 for a reliable fit, we wonder be using a 
plateau at $t=15-19$ for $t_{\rm src}=8$. 
Since the time slice $t=19$ or 20 is largely separated from the boundary at $t=T=32$, 
the boundary effect to the plateau as a whole
is expected to be small. Therefore $t_{\rm scr}=8$ is a reasonable choice. 
 
\begin{figure}[h]
\begin{center}
\includegraphics[width=100mm, angle=0] {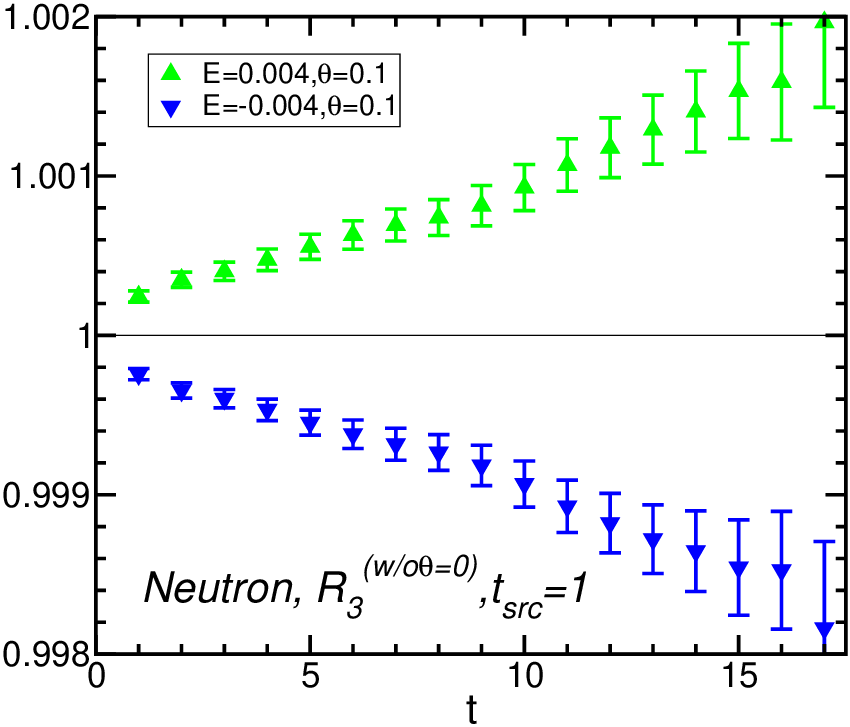}
\vskip 10mm
\includegraphics[width=100mm, angle=0] {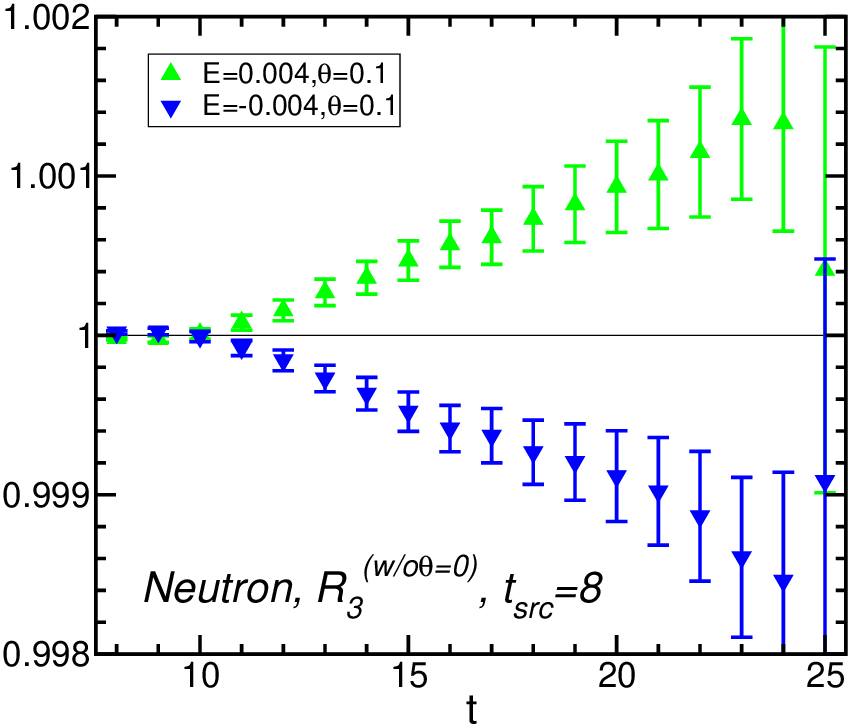}
\caption{These figures show the time dependence of $R_3^{(\rm w/o\theta=0)}(E,\theta,t)$ at
$\theta=0.1$ and $\kappa=0.1320$. Different figures show that source point $t_{\rm src}$ is 
in a different position in time direction. (top) $t_{\rm src}=1$ (bottom) $t_{\rm src}=8$. }
\label{fig:NEdn_clv_Ns24_src_corr}
\end{center}
\end{figure}

\begin{figure}[h]
\begin{center}
\includegraphics[width=100mm, angle=0] {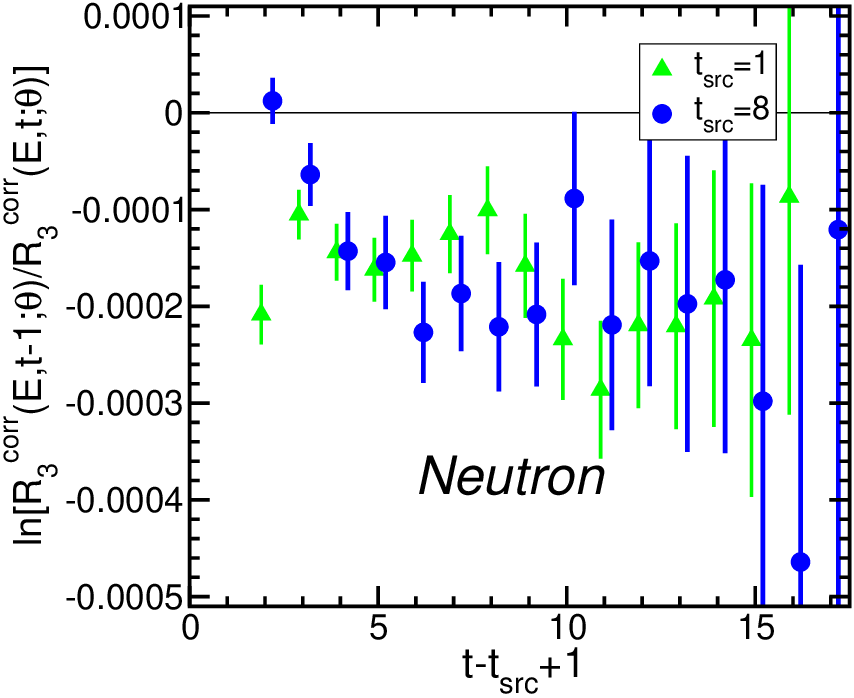}
\vskip 10mm
\includegraphics[width=100mm, angle=0] {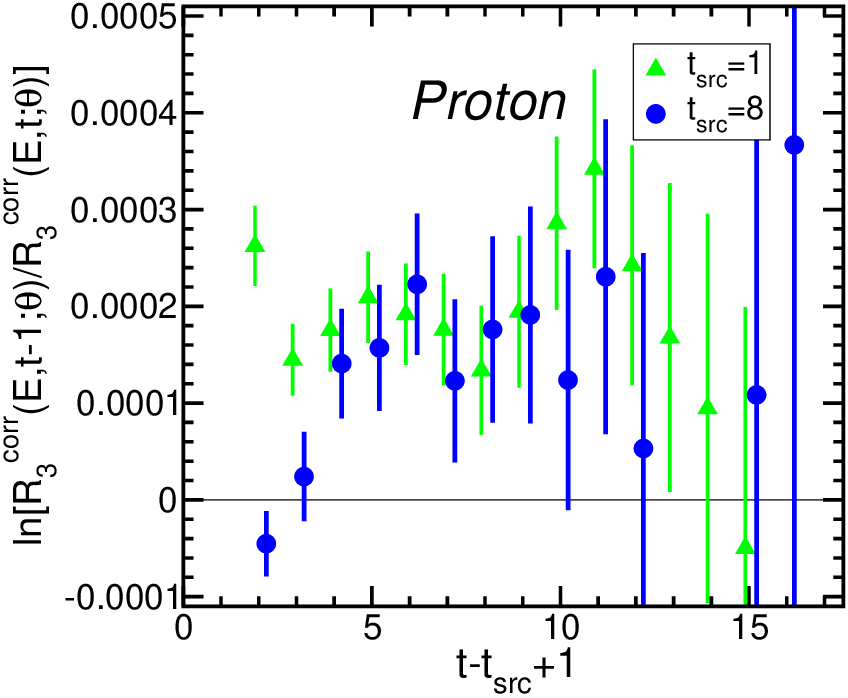}
\caption{These figures show the effective mass plot of the spin ratio, 
$R^{\rm corr}_3(E,t;\theta)$,
in both $t_{\rm src}=1$ and $t_{\rm src}=8$ at $\theta=0.1$ and $\kappa=0.1320$. }
\label{fig:NeffEdn_clv_Ns24_src_corr}
\end{center}
\end{figure}

In Fig.~\ref{fig:NEdn_clv_Ns24_src_corr} we compare the time dependence of 
$R_3^{(\rm w/o \theta=0)}(\pm E,\theta,t)$
for two different source points, $t_{\rm src}=1$ and $t_{\rm src}=8$.
We clearly observe a different time dependence of $R_3^{(\rm w/o \theta=0)}$ for two
sources at small time slices, $ t-t_{\rm src} \le 4$. 
We think that large deviations of
$R_3^{(\rm w/o \theta=0)}$ from unity at $t - t_{\rm src} \le 4$ for the $t_{\rm src}=1$ case
is an effect of the large non-uniform electric field near the boundary between $t=1$ and $t=T$.
On the other hand,  the deviation of $R_3^{(\rm w/o \theta=0)}$ from unity becomes visible 
around $t - t_{\rm src} \simeq 4$ for the case of $t_{\rm src}=8$. 
Since the plateau of the nucleon effective mass appears around  
$t - t_{\rm src} \simeq 5-6$,  
contributions from excited states to $R_3^{(\rm w/o \theta=0)}$ become small 
and the nucleon state dominates around this range of $t$ in the case of $t_{\rm src} = 8$.
In Fig.~\ref{fig:NeffEdn_clv_Ns24_src_corr} we plot the effective mass of 
$R_3^{\rm corr}(E,\theta=0,t)$ for the $t_{\rm src}=8$ case, together with
that for the  $t_{\rm src}=1$ case.
We notice that the plateau starts around $t-t_{\rm src} = 5$ for the
$t_{\rm src}=8$ case.
For the $t_{\rm src}=1$ case, on the other hand, the values of
effective mass of $R_3^{\rm corr}(E,\theta=0,t)$ around $t-t_{\rm src}
= 4-6$ seems smaller than the plateau of the $t_{\rm src}=8$ case,
suggesting that the boundary effects, observed in $R_3^{(\rm w/o
\theta=0)}$ at small $t-t_{\rm src}$,
still remain in the effective mass around $t-t_{\rm src} = 4-6$.
Therefore, to avoid possible contaminations from the boundary effect, 
we take sufficiently large separations such that $t-t_{\rm src}= 8-11$ 
for the fit of $R_3^{\rm corr}(E,\theta=0,t)$ in the case of $t_{\rm src}=1$.

An important lesson here is that we should take the starting point of the fitting range as far
from the source as possible, if the source is placed near the boundary such as
$t_{\rm src} =1$. This caution should be applied to all other data obtained with $t_{\rm src} =1$.

Fitting with $R_3^{\rm corr}$ exponentially in $5\le t-t_{\rm src}\le 9$ with $t_{\rm src}=8$, 
we obtain
\begin{equation}
  d_N = \left\{
\begin{array}{rc}
  -0.025(8) \,e\cdot\rm{fm} & \textrm{(Neutron)} \\
   0.024(11)\,e\cdot\rm{fm} & \textrm{(Proton)}  \\
\end{array}
\right. ,
\end{equation}
while for the $t_{\rm src}=1$ case we have
\begin{equation}
  d_N = \left\{
\begin{array}{rc}
  -0.030(8) \,e\cdot\rm{fm} & \textrm{(Neutron)} \\
   0.036(11)\,e\cdot\rm{fm} & \textrm{(Proton)}  \\
\end{array}
\right. 
\end{equation}
with $t-t_{\rm src}\in [7,11]$ as the fitting range.  
Two results are consistent with each other within large statistical errors.
Similarly, on a $16^3\times 32$ lattice, we obtain
\begin{equation}
  d_N = \left\{
\begin{array}{rc}
  -0.021(11) \,e\cdot\rm{fm} & \textrm{(Neutron)} \\
   0.026(13)\,e\cdot\rm{fm} & \textrm{(Proton)}  \\
\end{array}
\right. 
\end{equation}
for the clover fermion and
\begin{equation}
  d_N = \left\{
\begin{array}{rc}
  -0.017(8) \,e\cdot\rm{fm} & \textrm{(Neutron)} \\
   0.020(10)\,e\cdot\rm{fm} & \textrm{(Proton)}  \\
\end{array}
\right. 
\end{equation}
for the domain-wall fermion. The fitting range is $t-t_{\rm src}\in [6,11]$
with $t_{\rm src}=1$ for both fermions.
These values, summarized in Table \ref{tab:lat_EDM}, 
have the same sign and a similar order of magnitude
to the EDM form factor previously obtained on a $16^3\times 32$ lattice with the domain-wall fermion
with the form factor method, which is given by 
$F_3(q^2\simeq 0.58{\rm GeV}^2)/m_N=-0.024(5)$ e$\cdot$fm for neutron and 
$0.021(6)$ e$\cdot$fm for proton\cite{Shitani_form}. 
These agreements of sign and  magnitude between the two methods
support that the viability of this method explored in this paper.

\subsection{$E$ and $\theta$ dependence}\label{sec:EandTheta}
\begin{figure}[h]
\begin{center}
\vskip 10mm
\includegraphics[width=100mm, angle=0] {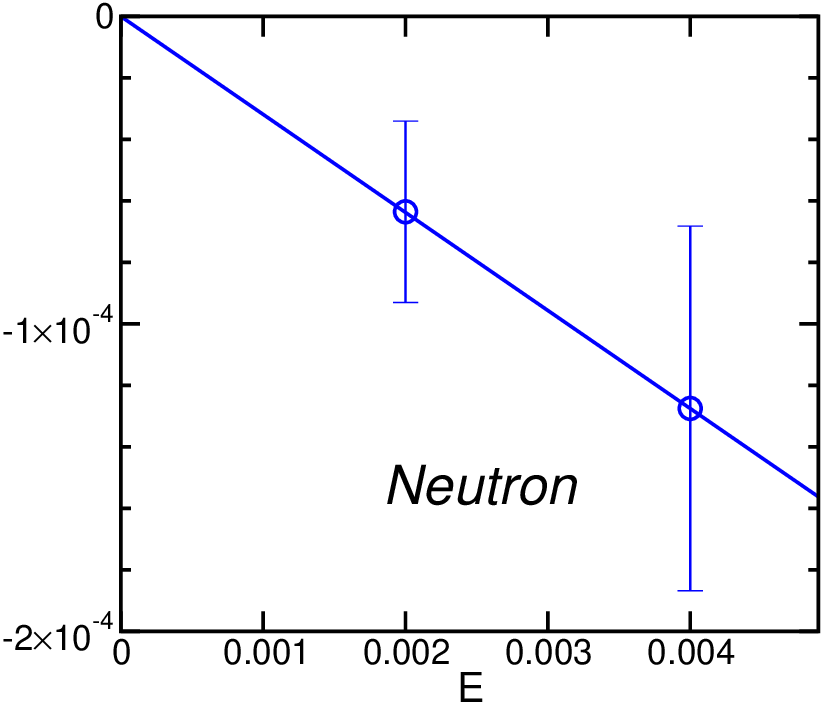}
\vskip 10mm
\includegraphics[width=100mm, angle=0] {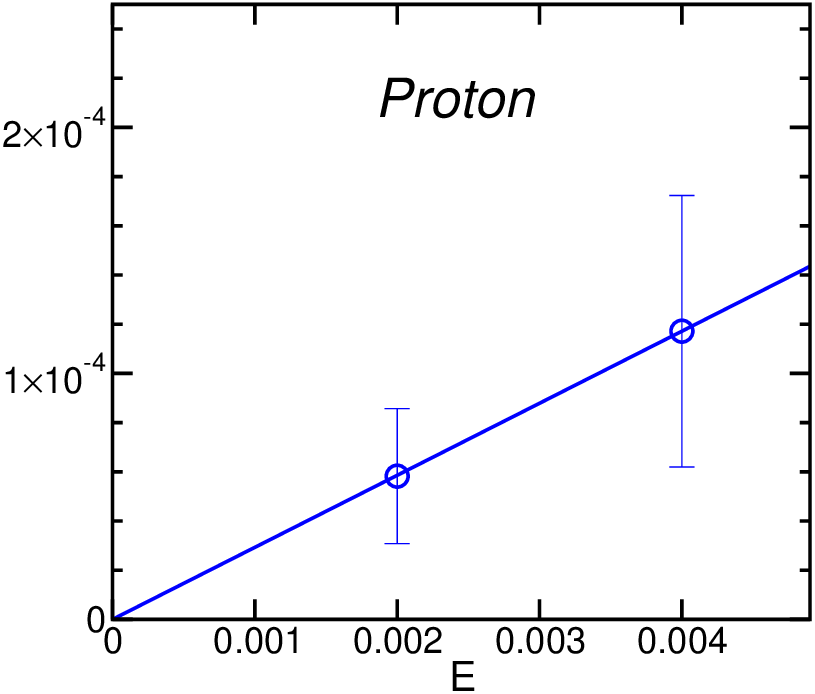}
\caption{The $E$ dependence for the fitting results of $R^{\rm corr}_3(E,t;\theta)$ 
in the range of $[7,12]$ at $\theta=0.1$ with domain-wall fermion  
(top) for neutron, (bottom) for proton. The solid line denotes a result of 
linear fit.}
\label{fig:Edn_Edep_DW}
\end{center}
\end{figure}

\begin{figure}[h]
\begin{center}
\vskip 10mm
\includegraphics[width=100mm, angle=0] {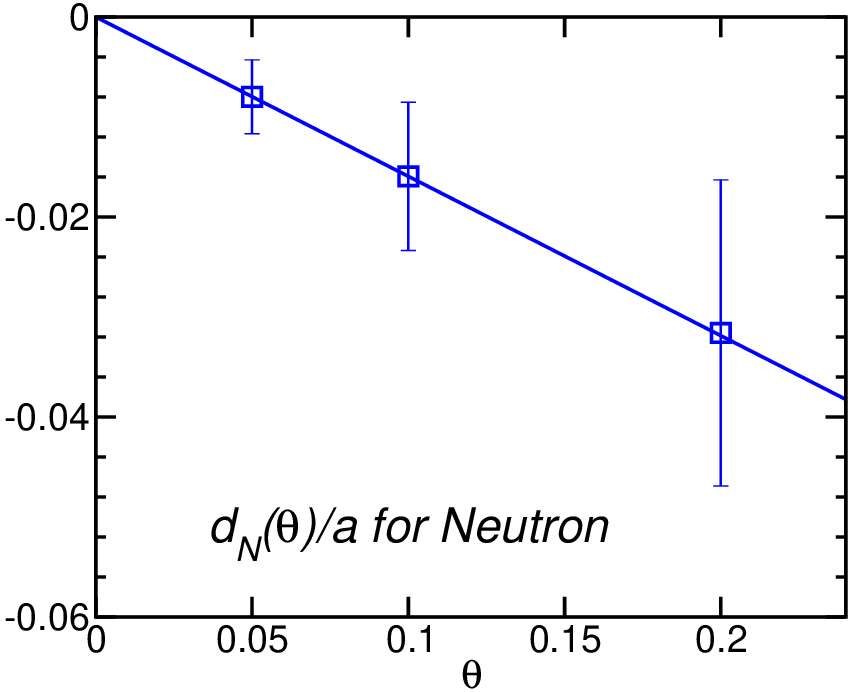}
\vskip 10mm
\includegraphics[width=100mm, angle=0] {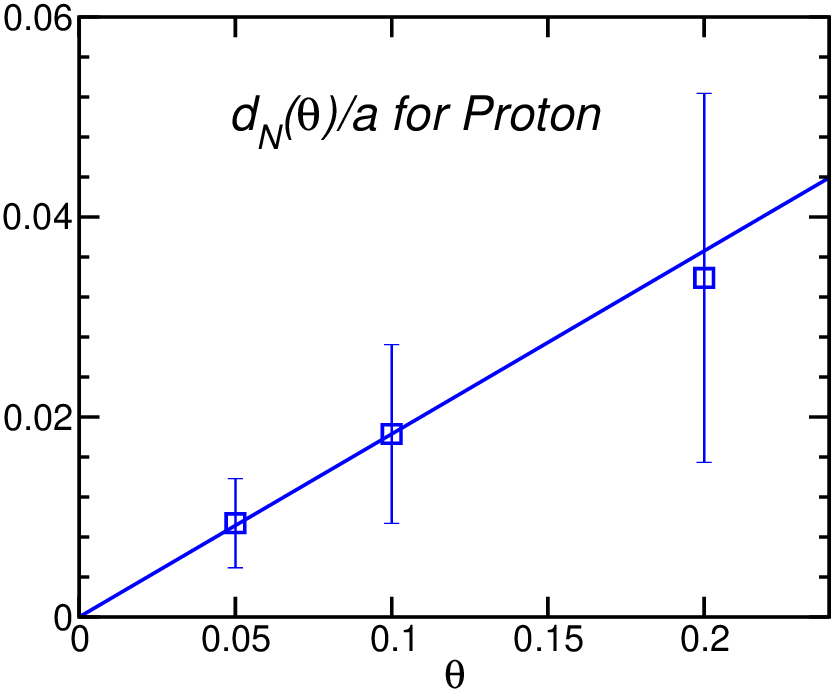}
\caption{The $\theta$ dependence for the fitting results of $R^{\rm corr}_3(E,t;\theta)$ 
in the range of $t\in [7,12]$ at $E=0.004$ with domain-wall fermion  
(top) for neutron, (bottom) for proton. The solid line denotes a result of 
linear fit.}
\label{fig:Edn_thetadep_DW}
\end{center}
\end{figure}

In Fig.~\ref{fig:Edn_Edep_DW} we plot values of EDM as a function of $E$ 
for neutron(upper) and proton (lower) at $\theta = 0.1$.
Observing the expected linear dependence on $E$ for both cases, we conclude that $O(E^3)$ 
contributions in (\ref{eq:EnergyDiff}) are negligible. 
Fig.~\ref{fig:Edn_thetadep_DW} shows $d_N(\theta)$ in lattice unit as a function of $\theta$ at
$E=0.004$, assuming the linear $E$ dependence of the fitted EDM signal.
We again confirm that the linearity in $\theta$ is good and thus $O(\theta^3)$ contributions
in (\ref{eq:small_theta}) are reasonably small.

We concluded that our choices of $(E, \theta)=(0.004,0.1)$ are small enough to
ensure linear dependences of the EDM signal on both $E$ and $\theta$, which we assume
in the analysis in the rest of this paper.

\subsection{Average over the electric field}
Averaging over three directions of the electric field
is not so useful in quenched simulation.
This way of increasing statistics, however,  may become important in full QCD case 
since the number of full QCD configurations is limited.
In this subsection we investigate the effectiveness of this method 
and the related question of 
the independence of the EDM signal on the direction of the electric field.

\begin{figure}[h]
\begin{center}
\vskip 10mm
\includegraphics[width=100mm, angle=0] {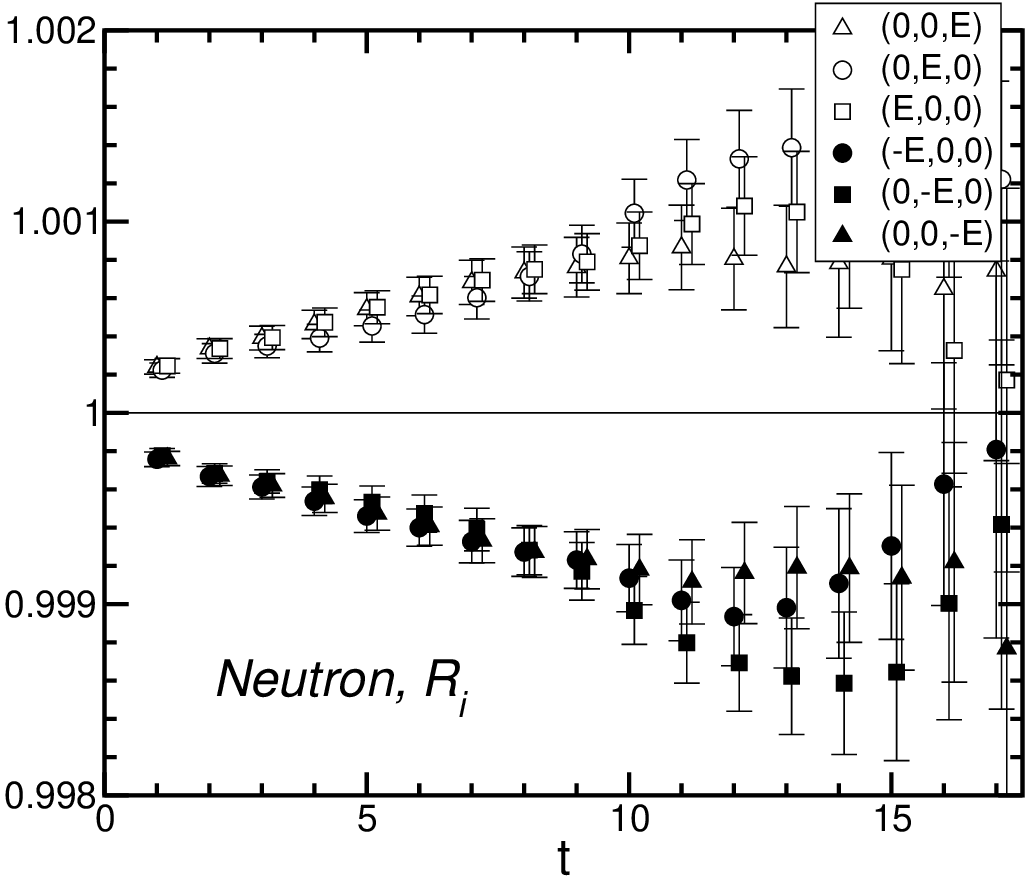}
\vskip 10mm
\includegraphics[width=100mm, angle=0] {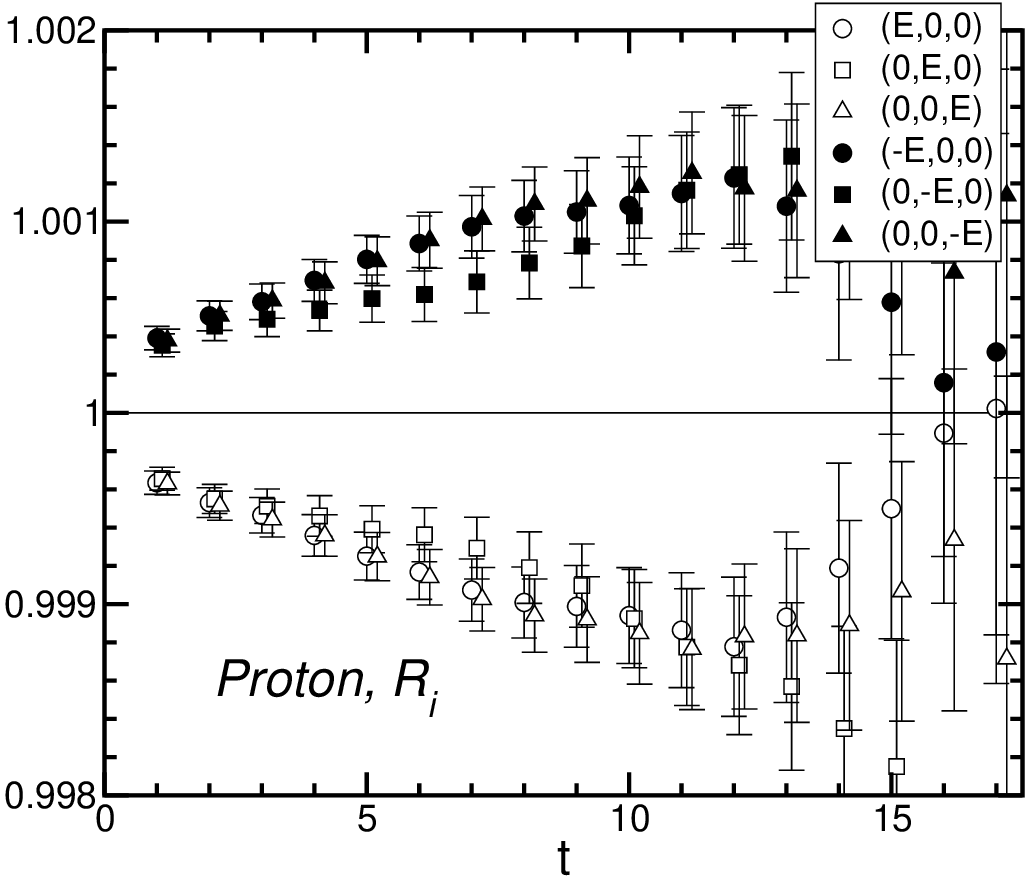}
\caption{The time dependence of $R_i(E,\theta,t)$ for 
three direction of electric field, which is $(E,0,0),\,(0,E,0),\,(0,0,E)$. 
The difference with open and filled symbol denote the different sign of electric 
field. (top) neutron case, (bottom) proton case. }
\label{fig:Edn_ratio_clv_Ns24}
\end{center}
\end{figure}

\begin{figure}[h]
\begin{center}
\includegraphics[width=100mm, angle=0] {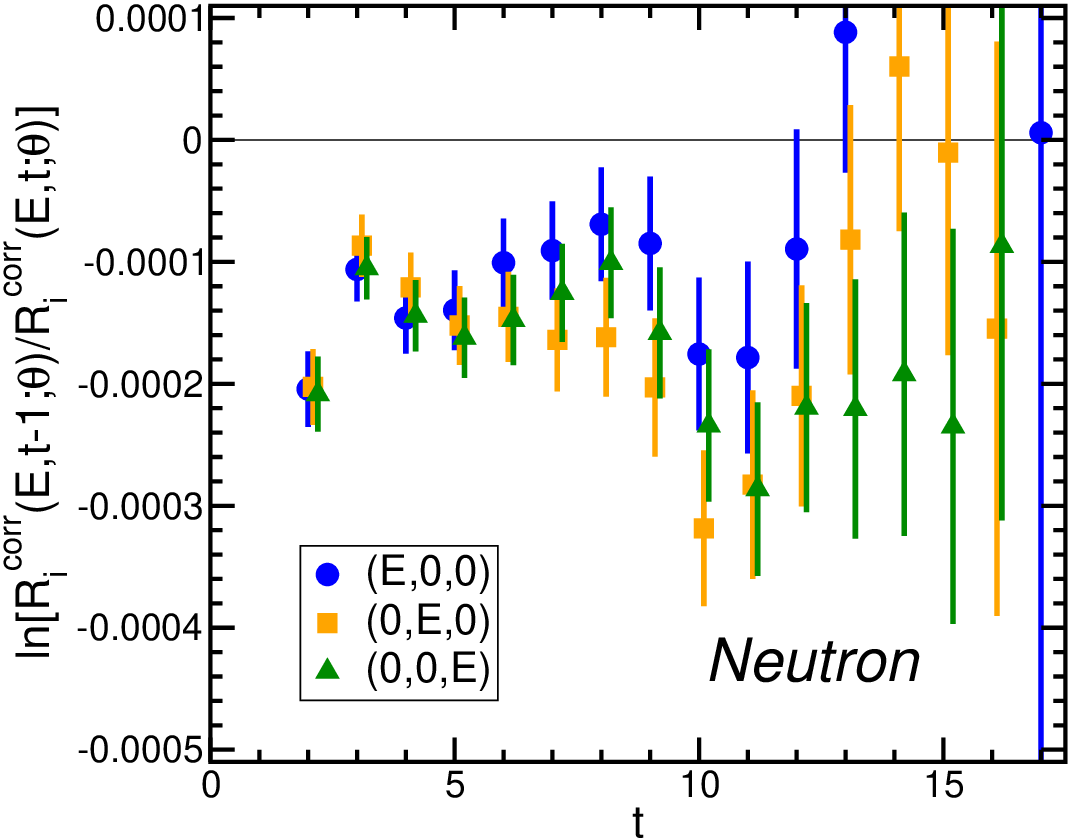}
\vskip 10mm
\includegraphics[width=100mm, angle=0] {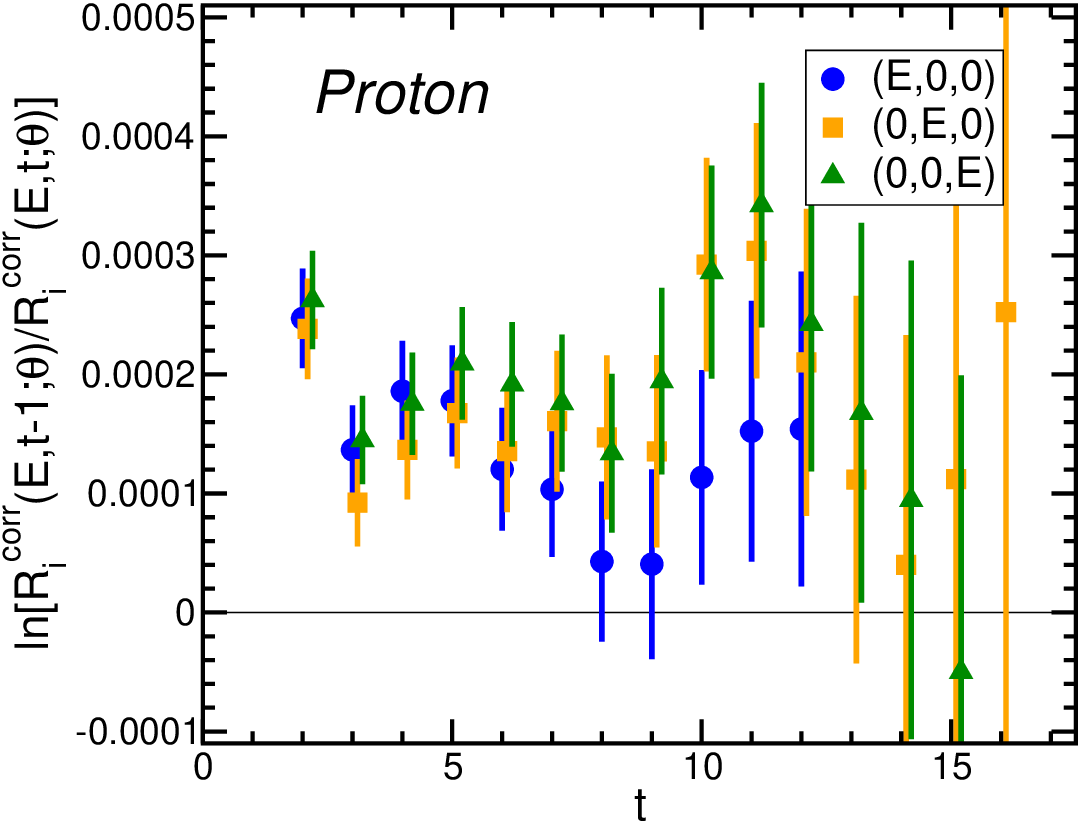}
\caption{The effective mass plot of $R^{\rm corr}_i(E,\theta,t)$ 
of each directions of electric field. (top) neutron case, (bottom) proton case. }
\label{fig:Edn_effratio_clv_Ns24_Dir}
\end{center}
\end{figure}

Using eq.~(\ref{eq:ratio_naive}), 
eq.~(\ref{eq:dn_x}) and eq.~(\ref{eq:dn_y}) for $\vec E=(0,0,E)$, $(E,0,0)$, 
and $(0,E,0)$, 
we obtain $R_i$ as a function of $E$ for each $\vec E$
on a $24^3\times 32$ lattice with the clover fermion at heaviest quark mass.
In Fig.~\ref{fig:Edn_ratio_clv_Ns24},
$R_i$ shows similar time dependences for all $i$ .
EDM signals, given in Fig.~\ref{fig:Edn_effratio_clv_Ns24_Dir}, 
are also comparable in the similar time range among different directions.
We confirm the consistency  among extraction of the EDM signal 
from three different directions using the
formulae in eqs.~(\ref{eq:ratio_naive})-(\ref{eq:dn_y}).  

\begin{figure}[h]
\begin{center}
\vskip 10mm
\includegraphics[width=100mm, angle=0] {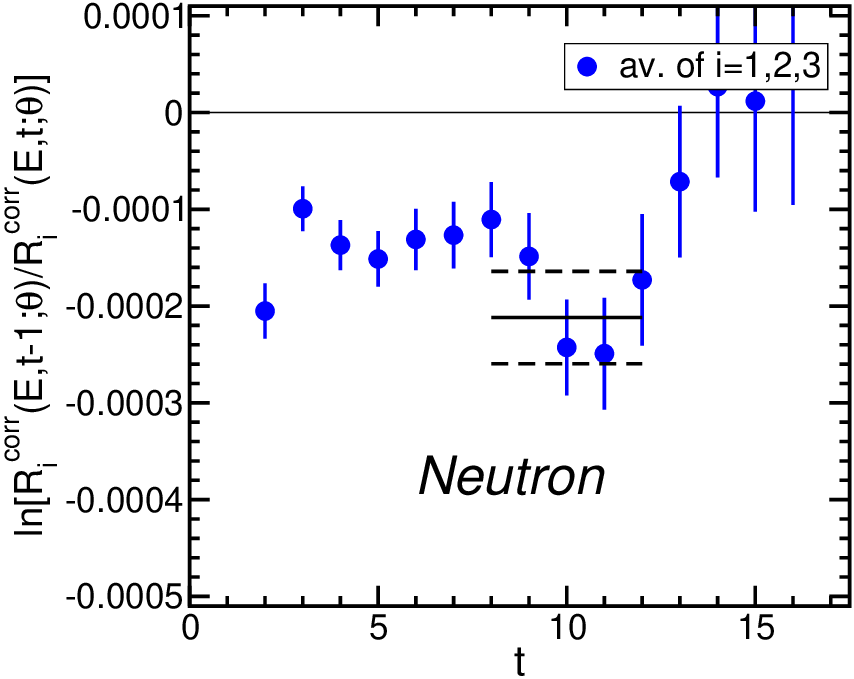}
\vskip 10mm
\includegraphics[width=100mm, angle=0] {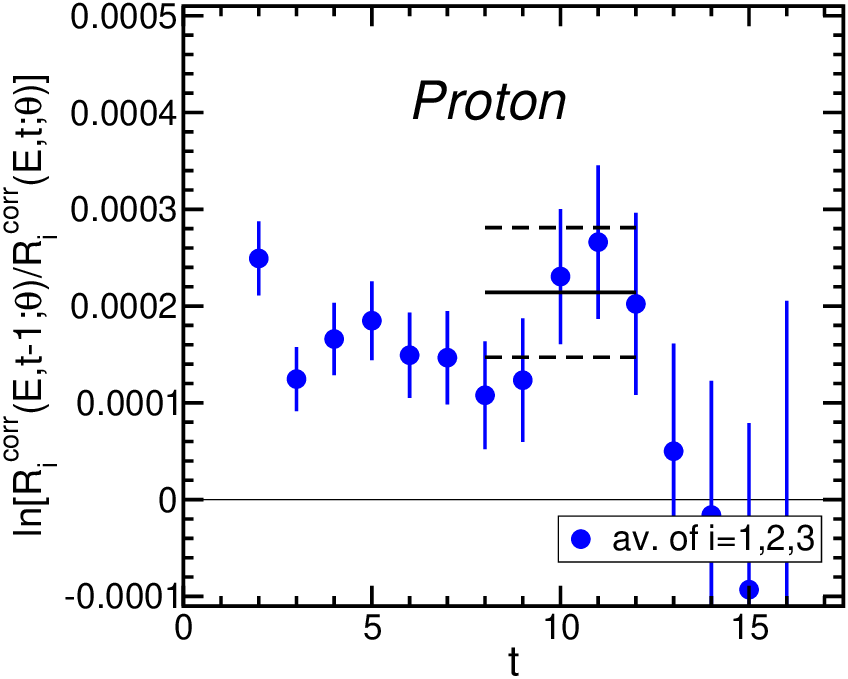}
\caption{These figures show the effective mass plot of the average of 
$R^{\rm corr}_i(E,\theta,t)$ for three directions of electric field with $\kappa=0.1320$. 
(top) neutron case, (bottom) proton case. }
\label{fig:Edn_effratio_clv_Ns24_Eav}
\end{center}
\end{figure}

We now consider the average over 3 directions. 
In Fig.~\ref{fig:Edn_effratio_clv_Ns24_Eav} the effective mass  
of the average, $R^{\rm corr} (E,\theta,t)\equiv \sum_i R_i^{\rm corr} (E,\theta,t)$
is plotted as a function of $t$.
Fitting it exponentially at $  7 \le t-t_{\rm src} \le 11$, we obtain
\begin{equation}
  d_N = \left\{
\begin{array}{rc}
  -0.0276(72)\,e\cdot\rm{fm} & \textrm{(Neutron)} \\
   0.0278(87)\,e\cdot\rm{fm} & \textrm{(Proton)} 
\end{array}
\right. .
\end{equation}
Although errors are reduced in the effective mass, 
the reduction in $d_N$ is much smaller than $1/\sqrt{3}$.
We conclude that the error reduction by this averaging  is limited, 
due to the possible correlation among $R_{i=1,2,3}(E,\theta,t)$,  


\section{Quark Mass Dependence}\label{sec:quark_mass_dep}
In this section we study the quark mass dependence of EDM 
using the clover fermion on a $24^3\times 32$ lattice.

\subsection{Quenched effects}\label{sec:quenched_eff}
It is well known in full QCD that EDM generated by the $\theta$ term
must vanish in the chiral limit. This can be seen from the fact that
the CP-violation Lagrangian after an appropriate chiral rotation \cite{Crewther}, 
\begin{equation}
  \delta\mathcal L_{\rm CP} = i\theta\bar m
  \sum_{i=u,d,s}\bar\psi_i\gamma_5\psi_i(x) ,\quad 
  \bar m = \Big(\sum_{i=u,d,s}m_i^{-1}\Big)^{-1},
\end{equation}
vanishes in the massless limit of any quarks. 
(See \cite{RBC} for more detailed argument on this property.)

In quenched QCD, however, this argument fails since
the $\theta$ parameter can not be translated to the above form
in the absence of the chiral anomaly, which requires the quark
determinant.
Therefore CP-violating observables generated by the $\theta$ term may remain non-zero
in the zero quark mass limit. 
Indeed, as discussed in \cite{RBC}, zero modes of the quark Dirac
operator can generate CP-odd contributions even in the massless limit.
It is not so easy, however, to determine the explicit quark mass dependence
of the EDM from the general argument in quenched QCD.

Recently, from the numerical simulation of the instanton liquid model \cite{Faccioli},
the $1/m_q^2$ dependence for NEDM has been reported near the chiral limit of quenched QCD.
The partially quenched chiral perturbation theory \cite{Connell}, on the other hand, 
has suggested  the $1/m_{\pi}^3$ behavior in the finite volume of $L^3$ 
at fixed sea quark mass $m_{\rm sea}$ such that
\begin{equation}
  d_N^{\rm P.Q.ChPT} \sim -\frac{e\theta m_{\rm sea}}{m_\pi^3L^3}f_\pi ,
\end{equation}
from the leading contribution of one-loop graphs.

\subsection{Quark mass dependence of EDM}
\begin{figure}[h]
\begin{center}
\includegraphics[width=100mm, angle=0] {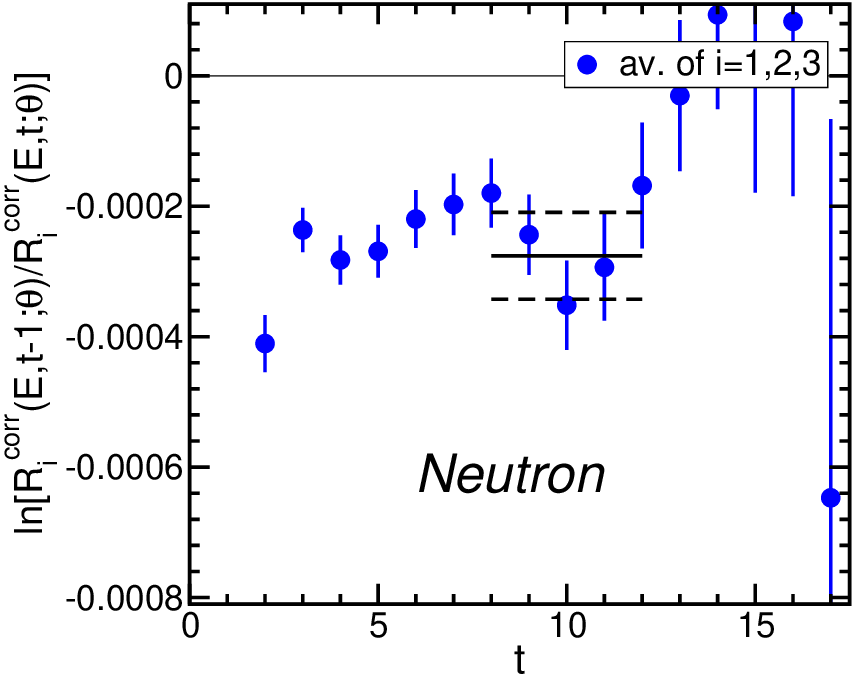}
\vskip 10mm
\includegraphics[width=100mm, angle=0] {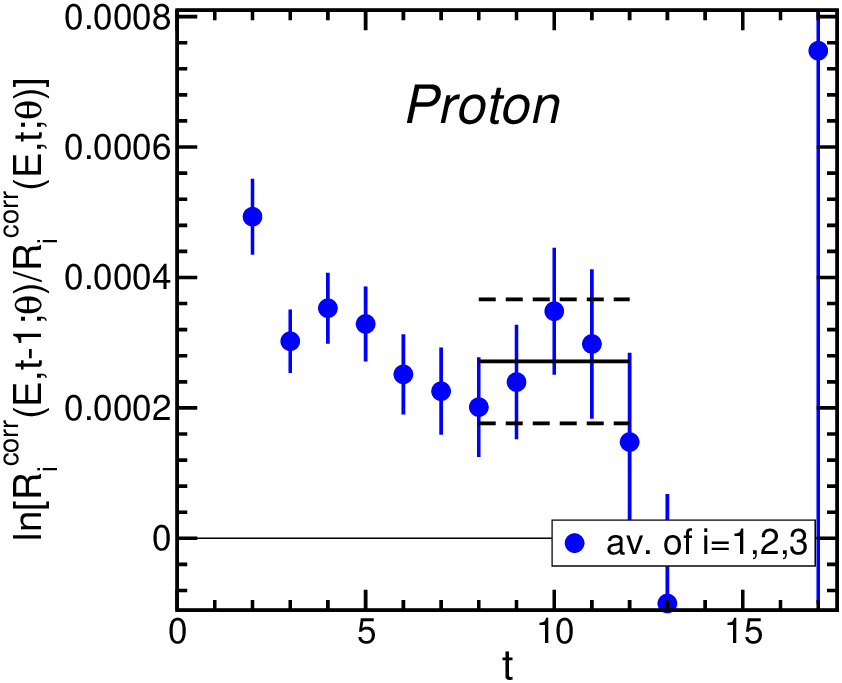}
\caption{The same figure as Fig.~\ref{fig:Edn_effratio_clv_Ns24_Eav} with $\kappa=0.1330$.
(top) neutron case, (bottom) proton case. }
\label{fig:Edn_effratio_clv_Ns24_Eav_K0133}
\end{center}
\end{figure}

\begin{figure}[h]
\begin{center}
\vskip 10mm
\includegraphics[width=100mm, angle=0] {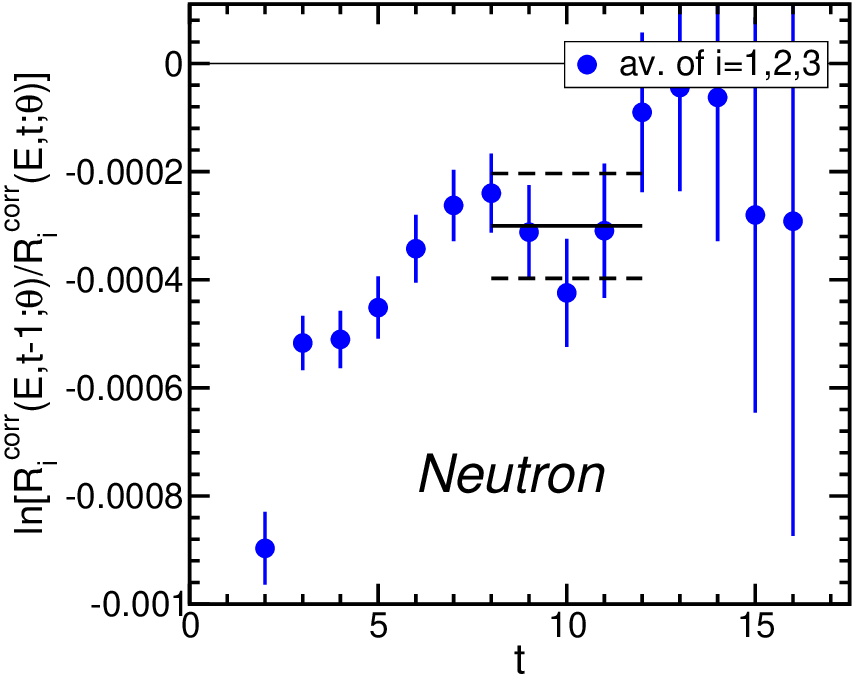}
\vskip 10mm
\includegraphics[width=100mm, angle=0] {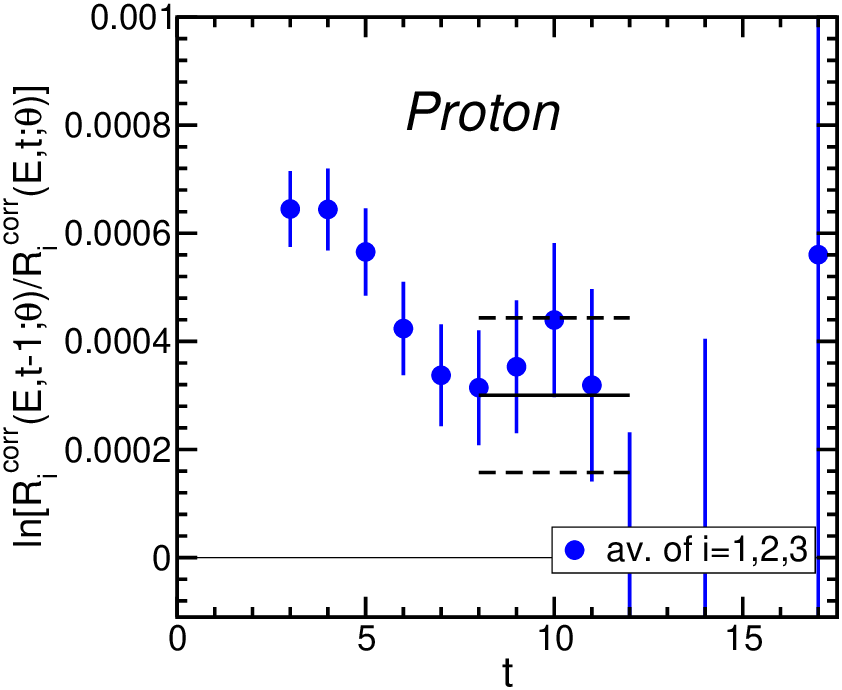}
\caption{The same figure as Fig.~\ref{fig:Edn_effratio_clv_Ns24_Eav} with $\kappa=0.1340$.
(top) neutron case, (bottom) proton case. }
\label{fig:Edn_effratio_clv_Ns24_Eav_K0134}
\end{center}
\end{figure}

\begin{figure}[h]
\begin{center}
\includegraphics[width=100mm, angle=0] {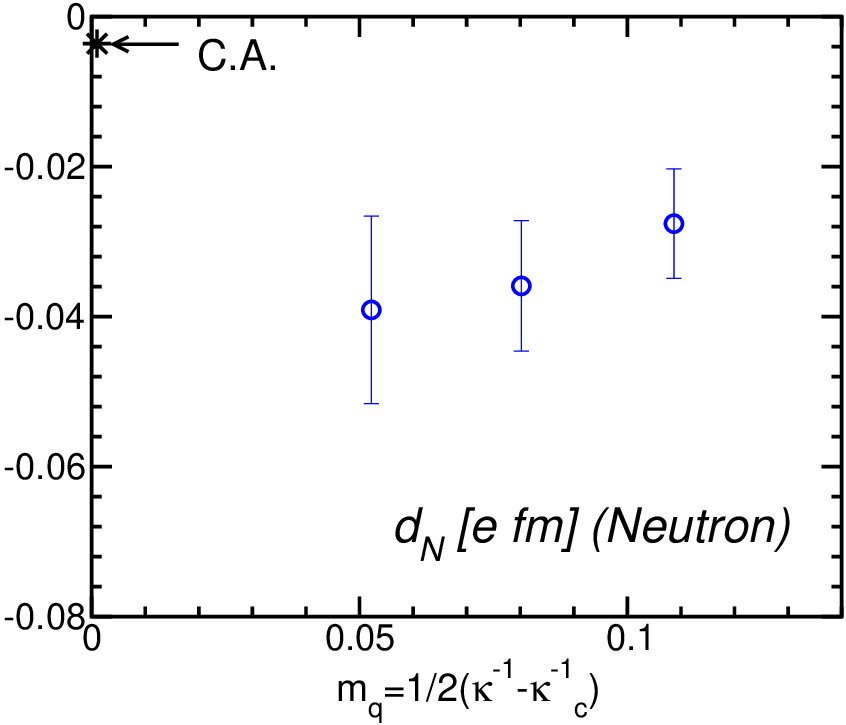}
\vskip 10mm
\includegraphics[width=100mm, angle=0] {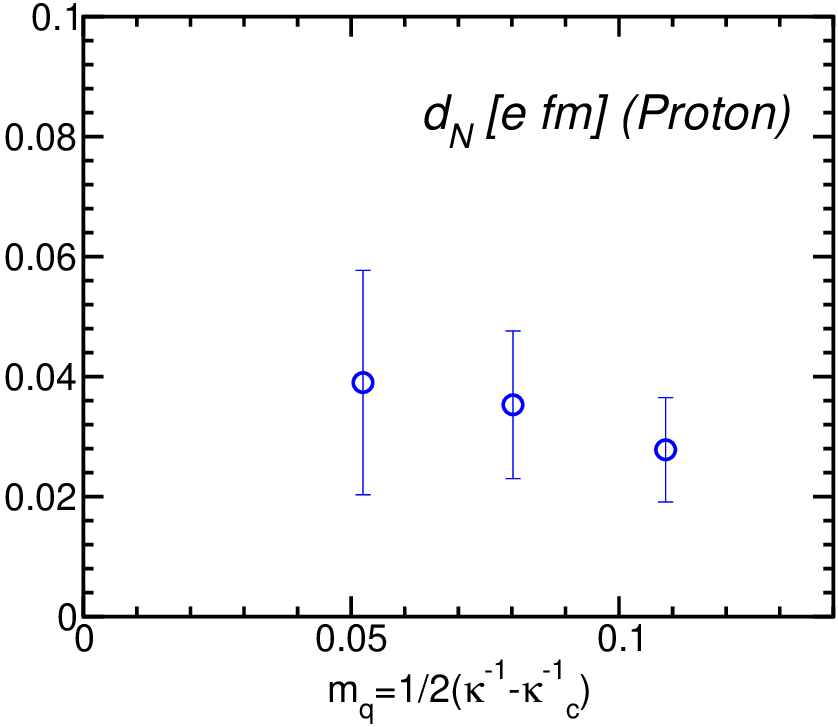}
\caption{The mass dependence of EDM factor with clover fermion. 
In top figure the star symbol shows the prediction from current algebra 
in \cite{Crewther}. 
(top) neutron case, (bottom) proton case. }
\label{fig:Edn_mdep}
\end{center}
\end{figure}

We calculate EDM at three different quark masses with the clover fermion
on a $24^3\times 32$ lattice.
In Fig.~\ref{fig:Edn_effratio_clv_Ns24_Eav_K0133} and 
Fig.~\ref{fig:Edn_effratio_clv_Ns24_Eav_K0134} we plot the effective mass of 
$R^{\rm corr}(E,\theta,t)=\sum_{i=1}^3 R^{\rm corr}_i(E,\theta,t)$ as a function of $t$
at two lighter quark mass with $t_{\rm src}=1$. 
Signals become a little noisier and less stable as the quark mass decreases.
Fitting data  at $t-t_{\rm src}\in [7,10]$ for the three quark masses,
we obtain the quark mass dependence of EDM for neutron and proton 
as shown in Fig.~\ref{fig:Edn_mdep} and Table~\ref{tab:massdep}. 
Compared with the current algebra result, $-0.0036$ e$\cdot$fm \cite{Crewther,Vecchia}
also shown in the top of Fig.~\ref{fig:Edn_mdep},
our quenched NEDM are about 10 times larger.
Moreover our results suggest that EDM does not vanish in 
the chiral limit for both neutron and proton.
We consider that the larger value of NEDM 
we focus is partly due to the  quenched effect.
Because of large statistical errors,
we can not distinguish 
the functional form of the mass dependence of EDM, whether
it stays constant or diverges in the chiral limit.

\subsection{Quark mass dependence of the CP-odd phase factor}
In addition to the EDM,  using the clover fermion,
we calculate a simpler quantity $f_N^1$,
the CP-odd phase factor of the nucleon propagator, defined in Ref.\cite{Shitani_form}  as
\begin{eqnarray}\label{eq:Npp_theta1}
  \langle N(\vec{p},t) \bar N(\vec{p},0) Q\rangle  =
  \vert Z_N\vert^2 e^{-E_{N} t}\frac{ f_N^1m_{N}}{2E_{N}} \gamma_5. 
\end{eqnarray}
Since the CP-odd phase factor arises from CP-violation effects of the  $\theta$ term, 
$f_N^1$ would vanish in the chiral limit of full QCD.  
In quenched QCD, however, this quantity also may remain non-zero in chiral limit 
because of the same reason as the EDM. 

\begin{figure}[h]
\begin{center}
\includegraphics[width=155mm, angle=0] {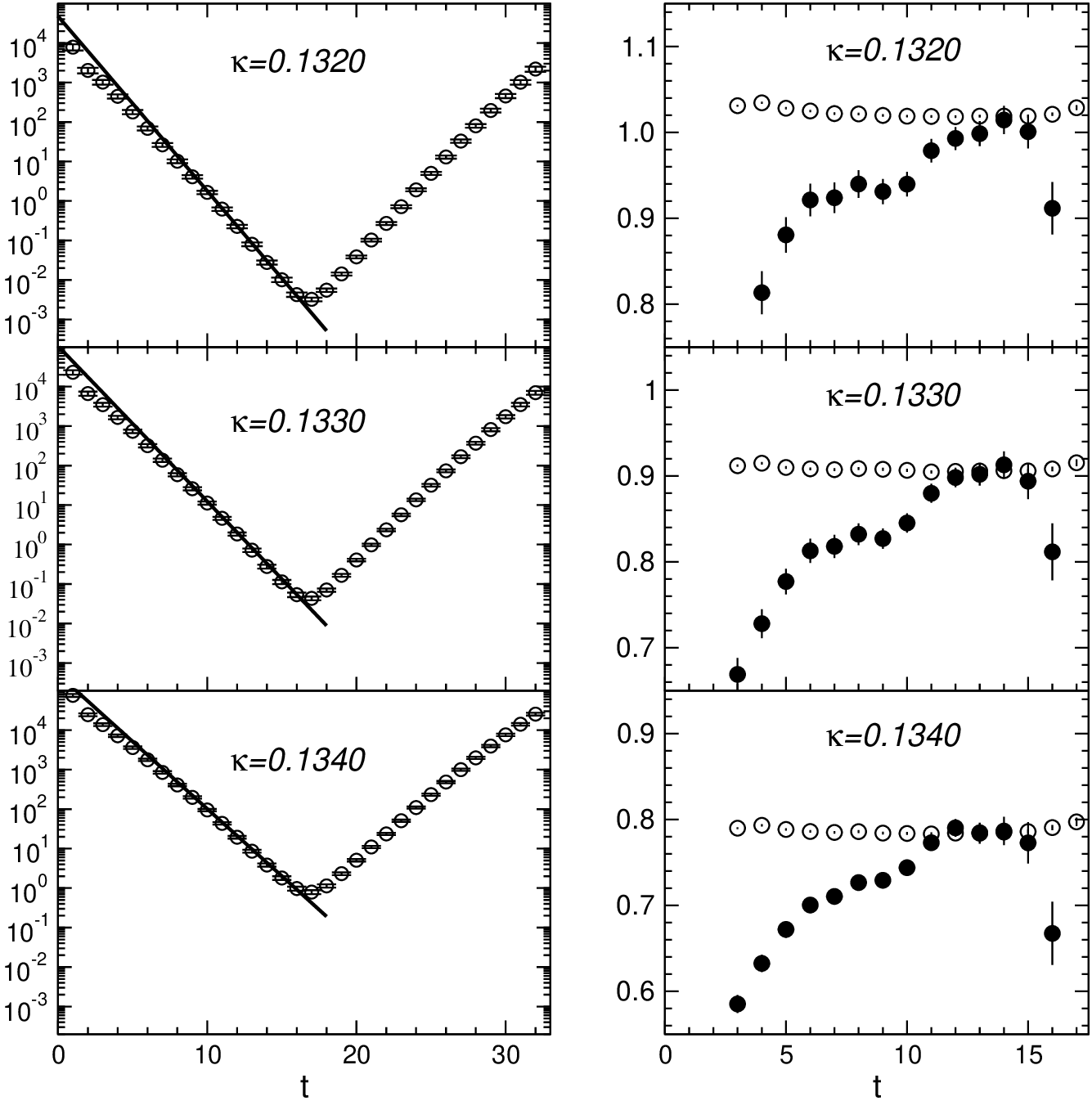}
\caption{In the left figure we show that the time dependence of nucleon propagator in 
next leading order of $\theta$ at each quark mass: $-{\rm tr}\langle NNQ\rangle \frac{\gamma_5}{2}$. 
The straight line represents fitting function with $f(x)=Ae^{-m_Nt}$.
In right figure we show that 
the comparison with exponent of nucleon propagator between the leading(open circles) 
and the next to leading order(solid circles).}
\label{fig:NEmass_Q}
\end{center}
\end{figure}

In Fig.~\ref{fig:NEmass_Q}, 
we show the time dependence of the nucleon propagator at the next leading in $\theta$, 
$-{\rm tr}[\langle N(\vec{0},t) \bar N(\vec{0},0) Q\rangle\frac{\gamma_5}{2}]$ (left),  
and effective masses of the leading nucleon propagator in $\theta$, 
${\rm tr}[\langle N(\vec{0},t) \bar N(\vec{0},0)\rangle\frac{1+\gamma_4}{2}]$,
as well as the next leading one (right) at 3 quark masses.
Since effective mass plots show the agreement of masses between two propagators
around $t=10$, we extract $f_N^1$ by fitting 
${\rm tr}[\langle N(\vec{0},t) \bar N(\vec{0},0) Q\rangle\frac{\gamma_5}{2}]$ 
at $9\le t \le 12$ in the form of (\ref{eq:Npp_theta1}), where $\vert Z_N\vert^2$ and $m_N$
have been fixed from the leading propagator.
 
\begin{figure}[h]
\begin{center}
\includegraphics[width=100mm, angle=0] {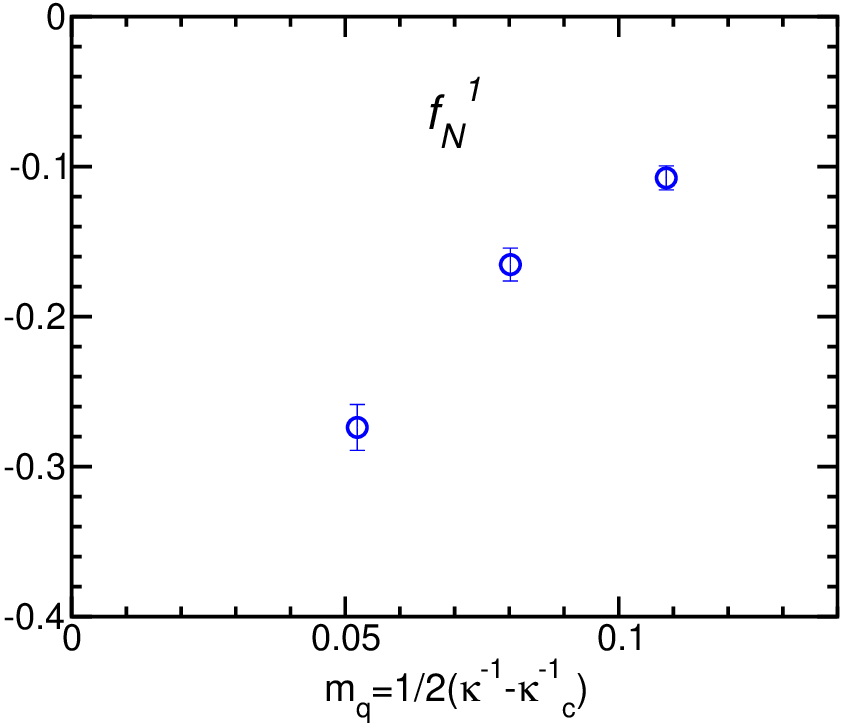}
\vskip 10mm
\includegraphics[width=100mm, angle=0] {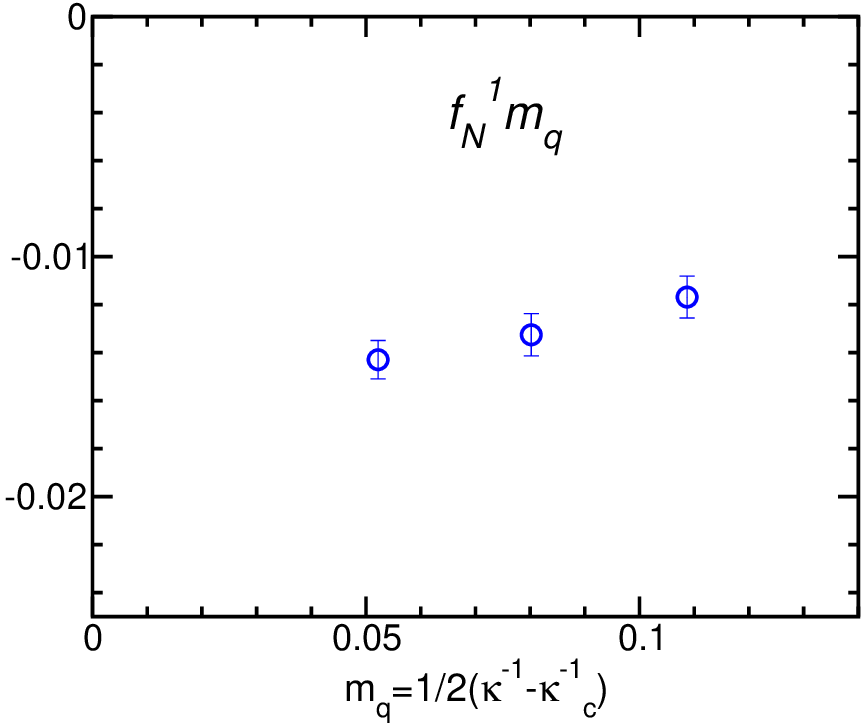}
\caption{The mass dependence of CP-odd phase factor with clover fermion.  
The top figure presents $f_N^1$ at each quark mass, and the bottom figure presents
the $f_N^1$ multiplied by quark mass.}
\label{fig:f1_mdep}
\end{center}
\newpage
\end{figure}

The quark mass dependence of $f_N^1$ is given in Fig.~\ref{fig:f1_mdep} and Table~\ref{tab:massdep}.
It is noted that errors of $f_N^1$ are much smaller than those of EDM.
The top of Fig.~\ref{fig:f1_mdep} shows that $f_N^1$ does not vanish in chiral limit
and moreover it seems to diverge as $1/m_q$ in this limit. 
To see this behavior more clearly, we plot $f_N^1$ multiplied by the quark mass 
$m_q=(\kappa^{-1}-\kappa_c^{-1})/2$ as a function of $m_q$ in the bottom of Fig.~\ref{fig:f1_mdep}. 
The fact that $f_N^1m_q$ seems almost constant at this range of the quark mass suggests 
that $f_N^1$ may diverge as $1/m_q$ in the chiral limit. 
It may be interesting to confirm this behavior of $f_N^1$ by some theoretical 
considerations.


\section{Summary and Discussion}\label{sec:summary}
In this paper, we have investigated the viability of an old idea for calculating the nucleon EDM
by introducing a uniform and static electric field. 
In this setup the nucleon EDM appears directly in the energy difference
between spin-up and spin-down states of the nucleon.
To introduce the complex $\theta$ term into lattice QCD calculations, we 
used the reweighting technique with the factor $e^{i\theta Q}$.
We have demonstrated that this reweighting method indeed works as long as $\theta$ is 
small enough, by calculating the nucleon EDM in quenched QCD on
a $16^3\times 32$ lattice at a relatively heavy quark mass.
We found that the quality of signals is not very sensitive to lattice fermion
formulations employed, domain-wall fermion and clover fermion in our study.
Using the clover fermion on a $24^3\times 32$ lattice,
we investigated the effect of non-uniformity of our electric field induced 
at the boundary in time direction. Even if the source point of nucleon is
placed near the boundary, the effect to the nucleon EDM disappears for large 
enough $t$, while the effect becomes smaller even at small $t$ if the source 
is placed away from the boundary.
We also found that the finite size effect to EDM is not so large: results between
(1.6 fm)$^3$ and (2.4 fm)$^3$ boxes agree within errors.

We investigated the quark mass dependence 
of the nucleon EDM and the CP-odd phase factor $f_N^1$ in quenched 
approximation on a larger volume with the clover fermion.
Both quantities do not seem to vanish in the chiral limit,
in contrast to full QCD where effects of the $\theta$ term 
disappear for a massless quark.
Therefore non-vanishing behaviors of EDM and $f_N^1$ are purely quenching effects. 
In particular, $f_N^1$ seems to diverge as $\mathcal O(1/m_q)$ in chiral limit. 
It is, however, difficult to determine precise quark mass dependences of 
these quantities in quenched QCD, due to larger statistical errors.

This work shows that the external electric field  method is simple and 
straightforward for the determination of the EDM in lattice QCD. 
In particular, the success with clover fermion in this method is significant
for applications to full QCD simulations.  
We are currently carrying out the EDM calculation using $N_f=2$ 
dynamical clover configurations generated by the CP-PACS collaboration\cite{Shintani_wE2}.

\section*{Acknowledgments}
This work is supported in part by Grant-in-Aid of the Ministry of Education
(Nos.
13135204, 
13135216, 
15540251, 
16540228, 
17340066, 
17540259, 
18104005, 
18540250  
).
In this work the numerical simulations have been carried out 
on the super parallel computer CP-PACS in University of Tsukuba and Hitachi SR11000  
in Hiroshima University and Tokyo University. 

\appendix
\section{Electric polarizability of Neutron}
In this appendix we discuss the electric polarizability of the neutron.
This observable can also be obtained by the external field method employed in
our calculation, as has been done in refs.~\cite{Fiebig,Christensen}. 
We compare our results with theirs.

\subsection{Definition}
The electric polarizability $\alpha_N$ is defined as the coefficient of the
$\vec E^2$ term in the expansion of the $\vec E$ dependent nucleon mass $m_N(\vec{E})$:
\begin{equation}\label{eq:def_pol}
  \Delta m_N(\vec E) = m_N(\vec E) - m_N(0) 
  = -\frac{1}{2}(4\pi\alpha_N)(e^{-1}a^{-2}\vec{E})^2 .
\end{equation}
which is measured by Compton scattering experiments.
Note that the electric field $\vec E$ here is dimensionless. 
A recent Compton scattering experiment gives 
\begin{equation}\label{eq:exp_pol}
  \alpha^{\rm exp}_N = (1.16\pm 0.15) \times 10^{-3}\,\textrm{fm}^3
\end{equation}
for the neutron\cite{PDG} .
In the lattice calculation the effective mass shift is calculated by
\begin{eqnarray}
  r_N(\vec E,t) &=& \frac{\langle N\bar N\rangle(\vec E,t)}{\langle
   N\bar N\rangle(\vec 0,t)},\\
  \Delta m_N(\vec E) &=& \ln\left[ \frac{r_N(\vec E,t)}{r_N(\vec E,t+1)}
			    \right].  \label{eq:mshift}
\end{eqnarray}
where $\langle N\bar N\rangle(\vec E,t)$ denotes the nucleon propagator
in the presence of the constant electric field $\vec E$ without reweighting $e^{i\theta Q}$. 
In order to remove spurious contributions odd in $\vec E$ from the effective mass shift, 
we take an average over $\vec E$ and $-\vec E$, by replacing
\begin{eqnarray}
r_N(\vec E) &\rightarrow&
  \frac{1}{2}({r_N(\vec E,t)}+{r_N(-\vec E,t)}) 
\end{eqnarray}
in eq.(\ref{eq:mshift}).

\subsection{Numerical results on a $16^3\times 32$ lattice } 
\label{sec:e_pol_PDB}
Our lattice setup for the calculation of the electric polarizability is same as
the one employed for the NEDM calculation in sec.~\ref{sec:EDM_clvDW}.
In particular,
the real electric field $\vec E=(0,0,E)$ in Minkovski space is 
introduced by the replacement of eq. (\ref{eq:inc_E}). Although the 
periodicity in time direction is broken by this electric field, the
boundary conditions for the fermion are periodic in both time and
spatial directions on a $16^3\times 32$ lattice. We employ the
domain-wall fermion at $E =4\times 10^{-3}$ and $E=2\times 10^{-3}$. 
As a comparison we also employ the clover fermion at $E=4\times 10^{-3}$.

\begin{figure}[h]
\begin{center}
\vskip 10mm
\includegraphics[width=100mm, angle=0] {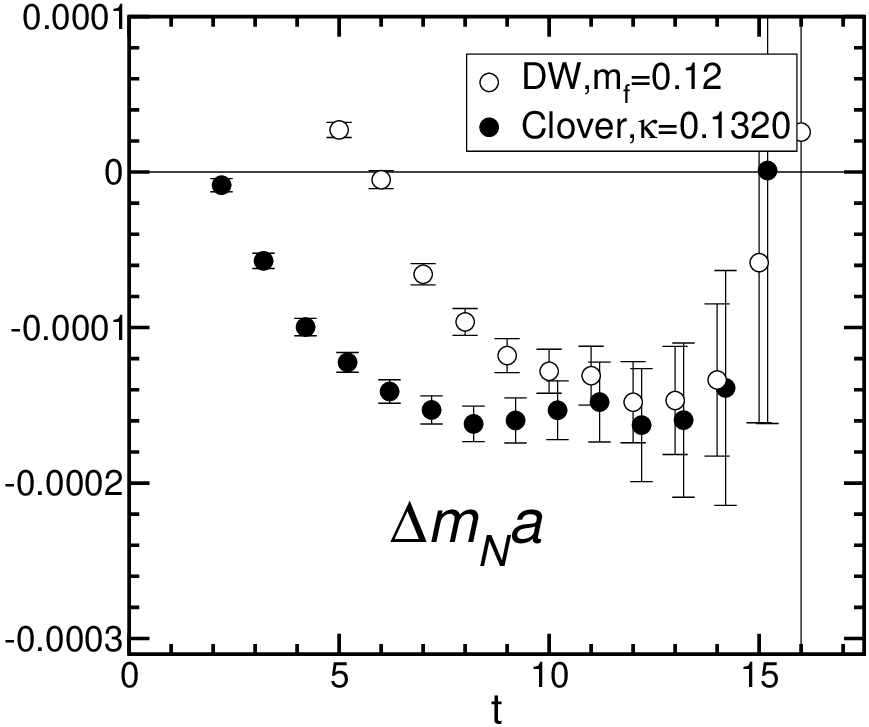}
\vskip 10mm
\includegraphics[width=100mm, angle=0] {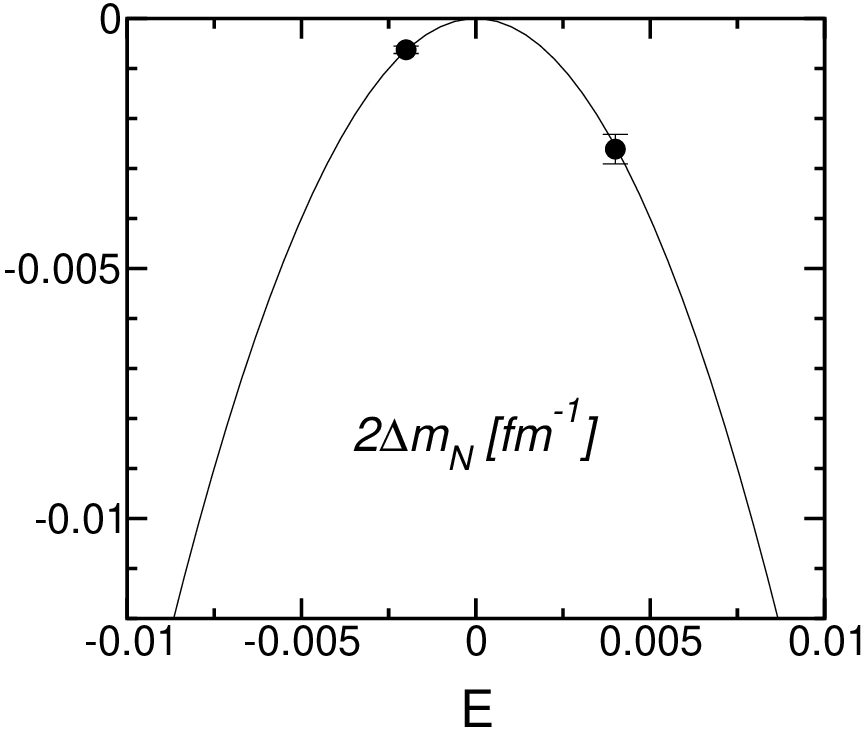}
\caption{(Top) The effective mass shift plot as Fig.~\ref{fig:mshift_Edep}
in periodic boundary condition with domain-wall and clover fermion 
at $E=4.0\times 10^{-3}$. 
(Bottom) The $E$ dependence of the mass shift of the above results 
from exponentially fitting $r_N$ in the range of $[9,14]$ with domain-wall fermion.
The solid line presents fit results with the function $f(E)=\alpha E^2$.}
\label{fig:mshift_Edep_DW}
\end{center}
\end{figure}

In the top of Fig.~\ref{fig:mshift_Edep_DW} we show the effective mass plot 
of $r_N$ in eq.~(\ref{eq:mshift}) for domain-wall and clover fermions on  
same configurations. 
We observe the plateau starting around $t=7$ for the clover fermion and
around $t=10$ for the domain-wall fermion.
From the exponential fit of $r_N(t)$ at $9\le t \le 14$, we obtain
$\Delta m_N$, whose values are given in Table~\ref{tab:mass_shift}.

In the bottom of Fig.~\ref{fig:mshift_Edep_DW} we present the $E$ dependence 
of the mass shift $2\Delta m_N$ for the domain-wall fermion.
By fitting data with $ -4\pi\alpha_N (e^2a^4)^{-1}E^2$, we obtain
the electric polarizability for neutron : 
\begin{equation}\label{eq:pol_period}
  \alpha_N = 1.32(2)\times 10^{-4}\,\textrm{fm}^{3}
\end{equation}
in the unit of $e^2(4\pi)^{-1}a^3\simeq 0.73\times 10^{-5}$ fm$^3$ with 
the fine-structure constant $\alpha = e^2/(4\pi) = 1/137$.

This value, obtained in quenched QCD at $a\simeq 0.1$ fm and $m_{PS}/m_V\simeq 0.88$.
is $1/10$ times smaller than the experimental value 
$\alpha_N^{\rm exp} = 1.16(15) \times 10^{-3}$ fm$^3$, 
but the sign of $\alpha_N$ agrees.

\subsection{Results on $24^3\times 32$ with two different source points}
We also calculate the electric polarizability of the neutron on a larger
volume, $24^3\times 32$, using the clover fermion at $\kappa=0.1320$. 
As in sec.~\ref{sec:bc_eff}, we employ two different source points,
$t_{\rm src}=1$ and $t_{\rm src}=8$, to investigate the effect of the
gap in $E$ at the boundary to the electric polarizability.

\begin{figure}[h]
\begin{center}
\vskip 10mm
\includegraphics[width=100mm, angle=0] {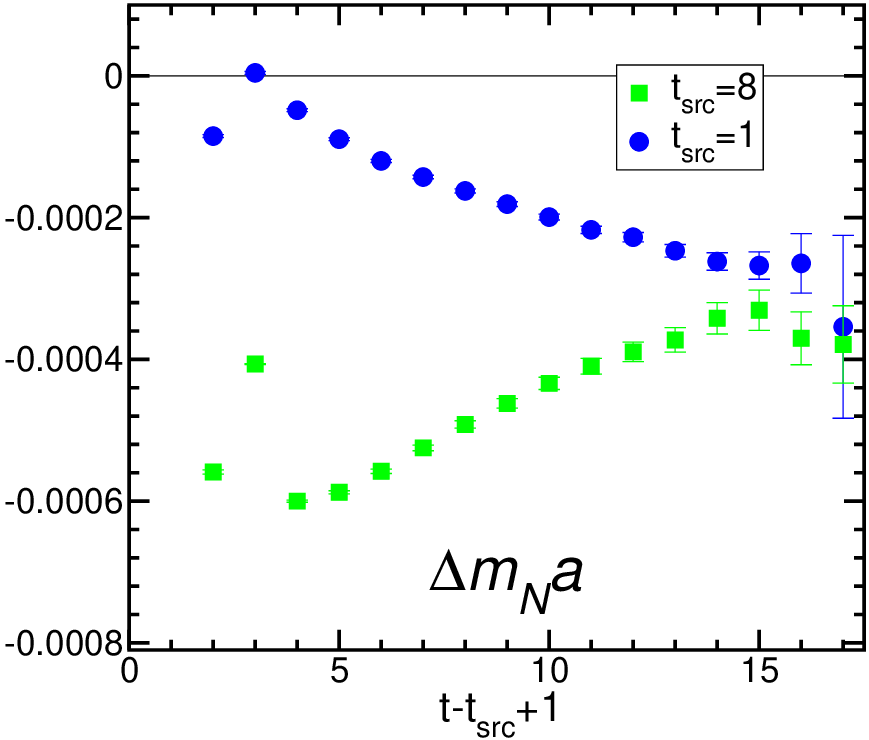}
\caption{This figure shows the comparison with different source point $t_{\rm src}$. 
We plot the effective mass shift as Fig.~\ref{fig:mshift_Edep}
in periodic boundary condition with clover fermion at $\kappa=0.1320$ 
and $E=4.0\times 10^{-3}$ in large lattice size $24^3\times 32$.}
\label{fig:Neffratio_Ns24clv}
\end{center}
\end{figure}

In Fig.~\ref{fig:Neffratio_Ns24clv} we present the effective mass shift, 
$\Delta m_Na$, for both $t_{\rm src}=1$ and $t_{\rm src}=8$. 
Compared with results on the smaller volume in sec.~\ref{sec:e_pol_PDB}, 
plateaus seem to appear at very large $t$ for both sources or even
$\Delta m_N a$ may not reach the plateau at $t\le 16$.
Even though an identification of plateaus is less reliable on the larger
volume, we fit data exponentially in $t$ at $13\le t \le 16$
and give values of $\Delta m_N a $ in Table \ref{tab:mass_shift}.
As seen in the table, the magnitude of fitted values is larger than the value
on the smaller volume.
We think that this discrepancy is mainly caused by contaminations from
excited states on the larger volume.
We need larger time separations to extract the ground state
contribution unambiguously.
We also observe large differences in the effective mass at small
$t$ between $t_{\rm src}=1$ and $t_{\rm src}=8$.
This indicates that the electric polarizability is quite sensitive to
the boundary effect. 

\begin{figure}[h]
\begin{center}
\vskip 10mm
\includegraphics[width=100mm, angle=0] {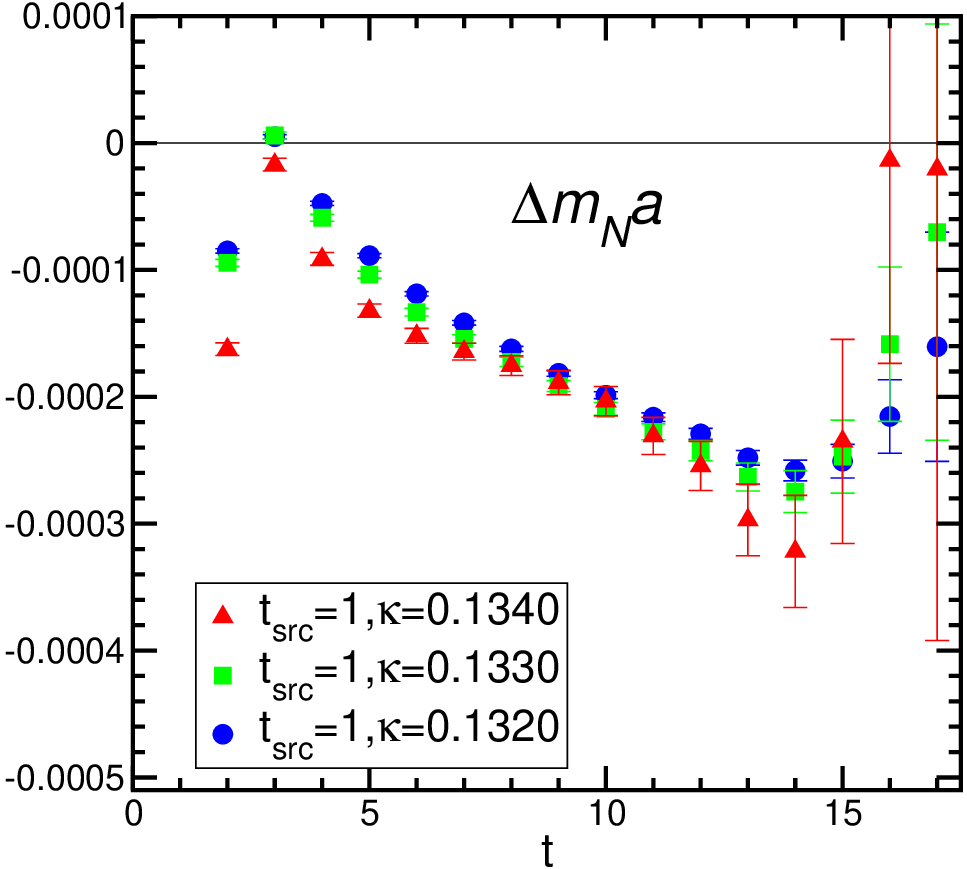}
\caption{We plot the effective mass shift as Fig.~\ref{fig:Neffratio_Ns24clv} with 
several quark masses.}
\label{fig:Neffratio_Eav_Ns24clv}
\end{center}
\end{figure}

\begin{figure}[h]
\begin{center}
\vskip 10mm
\includegraphics[width=100mm, angle=0] {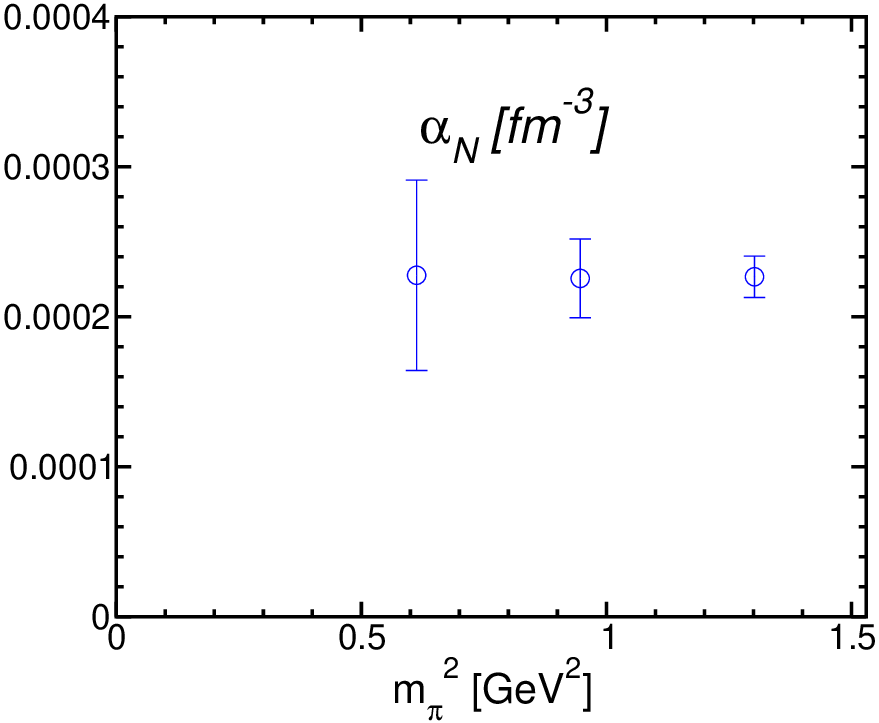}
\caption{This figure shows the mass dependence of electric polarizability of neutron 
with same parameter as Fig.~\ref{fig:Neffratio_Eav_Ns24clv}.}
\label{fig:Polmassdep_Ns24clv}
\end{center}
\pagebreak
\end{figure}

In Fig.~\ref{fig:Neffratio_Eav_Ns24clv} we plot the effective mass shift at different quark mass
after taken average over three directions of electric field with $t_{\rm src}=1$ 
on $24^3\times 32$. We observe that the time behavior is not so different with each other, 
and therefore its value will not depend on the quark mass strongly. 
Fig.~\ref{fig:Polmassdep_Ns24clv} and Table~\ref{tab:mass_shift_mdep}
shows the converted results to electric polarizability 
using fitting data of $\Delta m_N$ in each $\kappa$. 
In these heavier masses, the results seems to be constant for square of pion mass, 
though statistic errors are still large. Therefore 
more statistics are probably needed to give a precise value of
the neutron electric polarizability in the chiral limit.

\subsection{Comparison with previous calculations}

As a test of our method, we use same lattice parameters as in previous 
calculations\cite{Fiebig,Christensen}:
Accumulating 40 quenched configurations generated by the plaquette
action at $\beta=6.0$ ( $a\simeq 0.1$ fm ) on a $24^4$ lattice,
we calculate the electric polarizability by the Wilson fermion action
at $\kappa=0.1515$, which is the heaviest quark mass in \cite{Christensen}.
With the periodic boundary condition in spatial directions but
the Dirichlet boundary condition in the time direction,
the nucleon propagator is calculated for a point source at $t=1$ and 
a point sink at $t$. 

The electric field is introduced into all spatial link variables in the 
expanded form:
\begin{equation}\label{eq:def_im_E}
  U_3(x) \longrightarrow e^{iqEt}U_3(x)\simeq (1+iqEt)U_3(x) ,
\end{equation}
where we use an electric field in Euclidean space, which corresponds to 
the imaginary value in Minkovski space. 
Therefore the $E$ dependence of the mass shift $\Delta m_N$ is given by
\begin{equation}
  \Delta m_N(i\vec E) = -\frac{1}{2}(4\pi\alpha_N)(ie^{-1}a^{-2}\vec E)^2 
  = \frac{1}{2}(4\pi\alpha_N)e^{-2}a^{-4}\vec{E}^2
\end{equation}
with the electric polarizability $\alpha_N$.
As in \cite{Christensen}, we employ
$E=\pm 1.08\times 10^{-3},\,\pm 4.32\times 10^{-3},\,\pm 8.64\times 10^{-3}$
in the actual calculation. Note that the periodicity of spatial link
variables in the time direction is explicitly violated partly due to the fact
that $E\not= 2\pi/L$ and partly due to the expansion (\ref{eq:def_im_E}).

\begin{figure}[h]
\begin{center}
\vskip 10mm
\includegraphics[width=100mm, angle=0] {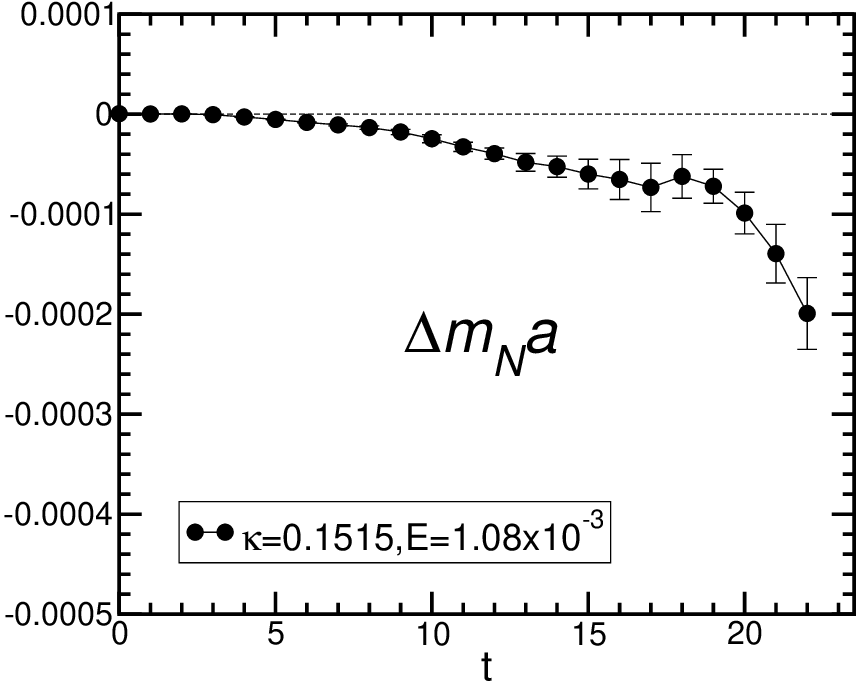}
\caption{The effective mass shift plot for neutron with Wilson fermion at 
$E=1.08\times 10^{-3}$ and $\kappa=0.1515$,
The initial time of our data is set up to be 
same as \cite{Christensen}.}
\label{fig:mshift}
\end{center}
\end{figure}

Fig.~\ref{fig:mshift} shows the effective mass shift for neutron 
in eq.~(\ref{eq:mshift}) at $|E|=1.08\times 10^{-3}$. 
Our data in Fig.~\ref{fig:mshift} roughly agree with filled circle symbols 
in Fig. 6 of \cite{Christensen}.  Unfortunately a candidate for a
possible plateau appears only at $15\le t\le 19$. Assuming that this is 
indeed a real plateau, we fit $\Delta m_N$ exponentially in $t$ at
$15\le t\le 19$ and gives values at each $E$ in  
Table~\ref{tab:mass_shift}. 

\begin{figure}[h]
\begin{center}
\vskip 10mm
\includegraphics[width=100mm, angle=0] {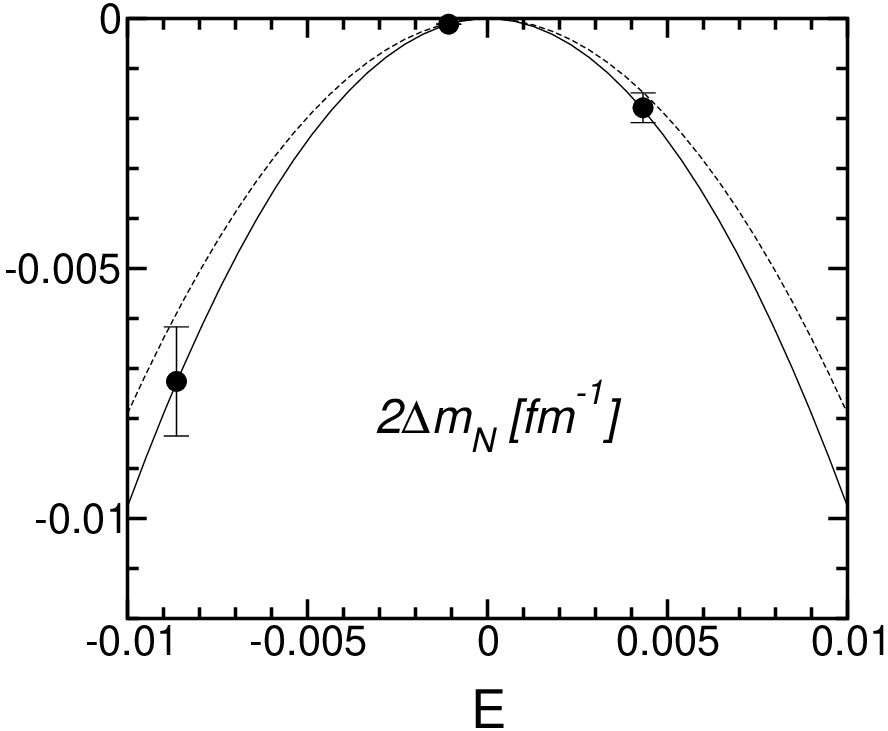}
\caption{This figure presents the $E$ dependence of the mass shift. 
The solid line denotes fitting results with our data and the broken line 
denotes results in \cite{Christensen}. Each our data are given by 
fitting in the time range of $[15,19]$.}
\label{fig:mshift_Edep}
\end{center}
\end{figure}

In Fig.~\ref{fig:mshift_Edep}  
we plot the $E$ dependence of mass shift $\Delta m_N$.
By fitting data with $\frac{1}{2}(4\pi\alpha_N) e^{-2}a^{-3} E^2$, 
we obtain a coefficient $\alpha_N$, the value of electric polarizability:  
\begin{equation}\label{eq:value_alpha}
  [\alpha_N]_{\rm Dirichlet} = -8.5(8)\times 10^{-4}\,\textrm{fm}^{3}.
\end{equation}
This value agree with the value in \cite{Christensen},
$[\alpha_N]_{\rm Dirichlet}=-7.9(5)\times 10^{-4}$ fm$^{3}$,
within about one-sigma error.
Surprisingly the sign of this result is opposite to the result 
(\ref{eq:pol_period}) obtained by the real electric field in Minkovski
space and to the experimental value in 
eq.~(\ref{eq:exp_pol})\footnote{
In \cite{Fiebig,Christensen} it has been claimed that the electric field
inserted as in eq.~(\ref{eq:def_pol}) is real so that their results of
the electric polarizability have the same sign as the experimental
value in eq.~(\ref{eq:exp_pol}). However, as shown here, the electric
field introduced by eq.~(\ref{eq:def_pol}) is real in Euclid space and 
it becomes pure imaginary in Minkovski space.
Therefore electric polarizabilities in \cite{Fiebig,Christensen} are 
opposite in sign to the experimental value. }.
In addition we confirm that the negative value of $\alpha_N$ is obtained 
even if we use the real electric field in Minkovski space in the Dirichlet
boundary condition. Therefore the wrong sign of $\alpha_N$ in this case
is not caused by the way of introducing the electric field 
(Euclid or Minkovski) but is related to the boundary condition in the
time direction. We think that $T=24$ is too short to suppress 
contributions from excited states to $\alpha_N$. 
In order to obtain a reliable estimate for $\alpha_N$, one should
investigate dependences of results on the lattice set-up such as the 
boundary conditions, the source point or the way of introducing the
electric field. We leave these studies in future investigations.

\vskip 5cm
\newpage

\begin{table}[t]
\begin{center}
\caption{Table for lattice parameters.
The column of $(A,B)$ denotes the smearing source parameter in the exponent}
\label{tab:lat_param}
\begin{tabular}{ccccccccc}
\hline
\hline
 Fermion  & $\beta$ & $L^3\times T\times N_s$ & $M$ & $a^{-1}\,[\textrm{GeV}]$ & $m_q$
    & $(A,B)$ & $m_{PS}/m_V$ & $m_Na$ \\
\hline
 Domain-wall & $2.6$ & $16^3\times 32\times 16$ & $1.8$ & $1.902(50)$ & $0.12$ 
    & (1.28,0.40) & $0.8781(4)$ & $1.1130(15)$ \\
\hline
\hline
 Fermion  & $\beta$ & $L^3\times T$ & $c_{SW}$ & $a^{-1}\,[\textrm{GeV}]$ & $\kappa$ 
    & $(A,B)$ & $m_{PS}/m_V$ & $m_Na$ \\
\hline
 Clover & $2.6$ & $16^3\times 32$ & $1.340$ & $1.902(50)$ & $0.1320$ 
& (1.55,0.24) & $0.8508(5)$ & $1.0202(17)$ \\
\hline
 Clover & $2.6$ & $24^3\times 32$ & $1.340$ & $1.902(50)$ & $0.1320$ 
& (1.55,0.35) & $0.8494(1)$ & $1.0186(9)$ \\
        &       &                 &         &             & $0.1330$ 
& (1.55,0.31) & $0.8026(2)$ & $0.9058(14)$ \\
        &       &                 &         &             & $0.1340$ 
& (1.55,0.27) & $0.7253(2)$ & $0.7843(16)$ \\
        &       &                 &         &             & $\kappa_c=0.1359(1)$
& & & \\
\hline
\hline
\end{tabular}
\end{center}
\pagebreak
\end{table}
\begin{table}[t]
\begin{center}
\caption{Table for EDM results in some lattice parameters}
\label{tab:lat_EDM}
\begin{tabular}{ccccccccc}
\hline
\hline
  fermion & $m_Na$ & lattice size & source point & fitting range & $d_N$ (Neutron) & $d_N$ (Proton) \\
\hline
  domain-wall & $1.1130(15)$ & $16^3\times 32$ & $t_{\rm src}=1$ & $t-t_{\rm src}\in [6,11]$ 
    & $-0.0170(79)$ & $0.0196(95)$ \\
  clover      & $1.0202(17)$ & $16^3\times 32$ & $t_{\rm src}=1$ & $t-t_{\rm src}\in [6,11]$ 
    & $-0.0205(104)$ & $0.0256(125)$ \\
  clover      & $1.0186(9)$ & $24^3\times 32$ & $t_{\rm src}=1$ & $t-t_{\rm src}\in [7,11]$ 
    & $-0.0304(78)$ & $ 0.0361(111)$ \\
  clover      & $1.0200(9)$  & $24^3\times 32$ & $t_{\rm src}=8$ & $t-t_{\rm src}\in [5,9]$ 
    & $-0.0246(83)$ & $0.0237(112)$ \\
\hline
\hline
\end{tabular}
\end{center}
\pagebreak
\end{table}
\begin{table}[t]
\begin{center}
\caption{The mass dependence of EDM factor from exponential fit in the range 
$8\le t\le 12$ for 
$R^{\rm corr}(E,\theta,t)$ which is the average over indices $i=1,2,3$ and CP-odd phase factor 
in the next leading term of nucleon propagator.}
\label{tab:massdep}
\begin{tabular}{cccccccc}
\hline
\hline
 &  Neutron & & & Proton & & $f_N^1$ & $f_N^1m_q$\\
\hline
 $\kappa$ & fit & $d_N\,[\textrm{e}\cdot\textrm{fm}]$ & 
          & fit & $d_N\,[\textrm{e}\cdot\textrm{fm}]$ \\
\hline
 0.1320 & $-0.000212(48)$ & $-0.0276(72)$  & & $0.000214(67)$  & $0.0278(87)$ & $-0.1075(80)$ 
        & $-0.0117(8)$ \\
 0.1330 & $-0.000276(67)$ & $-0.0359(87)$ & & $0.000271(95)$  & $0.0353(123)$ & $-0.1653(111)$
        & $-0.0133(9)$ \\
 0.1340 & $-0.000300(97)$ & $-0.0391(125)$ & & $0.000300(143)$ & $0.0390(187)$ & $-0.2738(152)$
        & $-0.0143(8)$ \\
\hline
\hline
\end{tabular}
\end{center}
\pagebreak
\end{table}
\begin{table}[t]
\begin{center}
\caption{Summary of the fitting results of mass shift of neutron with different 
boundary condition and fermions.}
\label{tab:mass_shift}
\begin{tabular}{cccccccc}
\hline
\hline
  gauge action & mass & lattice size & B.C. & $t_{\rm src}$ & $E$ & $\Delta m_Na$ \\
\hline
 Domain-wall fermion & & & & & & \\
\hline
RG Iwasaki $\beta=2.6$ & $m_f=0.12$ & $16^3\times 32$ & Periodic & $t_{\rm src}=1$ 
            & Real, 0.002 & $-$0.0000375(44) \\
         &  &                 &          &                 & Real, 0.004 & $-$0.000157(18) \\
\hline
Clover fermion & & & & & & \\
\hline
RG Iwasaki $\beta=2.6$ & $\kappa=0.1320$ & $16^3\times 32$ & Periodic & $t_{\rm src}=1$ 
       & Real, 0.004 & $-$0.000155(20) \\
    &  & $24^3\times 32$ & Periodic  & $t_{\rm src}=1$ & Real, 0.004 & $-$0.000265(22) \\
    &  &                 &           & $t_{\rm src}=8$ & Real, 0.004 & $-$0.000356(50) \\
\hline
Wilson fermion & & & & & & \\
\hline
Plaquette $\beta=6.0$ & $\kappa=0.1515$ & $24^3\times 24$ & Dirichlet & $t_{\rm src}=1$ 
       & Imag, 0.00108 & $-$0.000069(2) \\
    &  &                 &           &                 & Imag, 0.00432 & $-$0.00107(18) \\
    &  &                 &           &                 & Imag, 0.00864 & $-$0.00435(65) \\
\hline
\hline
\end{tabular}
\end{center}
\pagebreak
\end{table}
\begin{table}[t]
\begin{center}
\caption{Summary of the fitting results of polarizability of neutron with clover fermion 
action at three different quark masses after average over three directions of electric field.}
\label{tab:mass_shift_mdep}
\begin{tabular}{cccccccc}
\hline
\hline
  gauge action & lattice size & B.C. & mass & $t_{\rm src}$ & $E$ & $\alpha_N\,({\rm fm}^{-3})$\\
\hline
\hline
RG Iwasaki $\beta=2.6$ & $24^3\times 32$ & Periodic & $\kappa=0.1320$ & $t_{\rm src}=1$ 
            & Real, 0.004 & 0.000227(14) \\
         &  &    & $\kappa=0.1330$ & $t_{\rm src}=1$ & Real, 0.004 & 0.000226(26) \\
         &  &    & $\kappa=0.1340$ & $t_{\rm src}=1$ & Real, 0.004 & 0.000228(63) \\
\hline
\hline
\end{tabular}
\end{center}
\end{table}
\end{document}